\documentclass[useAMS,usenatbib,usegraphicx]{mn2e}
\usepackage{txfonts}
\makeatletter
\DeclareMathAlphabet{\mathpzc}{OT1}{pzc}{m}{it}
\newcommand\cf{\mathpzc{F}}
\newcommand\h{\ensuremath{^\rmn h}}
\newcommand\m{\ensuremath{^\rmn m}}

\title[The Variable Stars in M31]
{The POINT-AGAPE Survey I: The Variable Stars in M31.
\thanks{Based on observations made through the Isaac Newton Group's
Wide Field Camera Survey Programme with the Isaac Newton Telescope
operated on the island of La Palma by the Isaac Newton Group in the
Spanish Observatorio del Roque de los Muchachos of the Instituto de
Astrofisica de Canarias.}}
\author[J.~An et al.]
{Jin~H.~An,$^1$\thanks{E-mail: jin@ast.cam.ac.uk}
N.~W.~Evans,$^1$
P.~Hewett,$^1$
P.~Baillon,$^2$
S.~Calchi~Novati,$^3$
B.~J.~Carr,$^4$\newauthor
M.~Cr\'ez\'e,$^{5,6}$
Y.~Giraud-H\'eraud,$^6$
A.~Gould,$^7$
Ph.~Jetzer,$^3$
J.~Kaplan,$^6$
E.~Kerins,$^8$\newauthor
S.~Paulin-Henriksson,$^6$
S.~J.~Smartt,$^1$
C.~S.~Stalin,$^6$
and Y.~Tsapras.$^4$\newauthor
(The POINT-AGAPE\thanks{Pixel-lensing Observations on the Isaac
Newton Telescope - Andromeda Galaxy Amplified Pixels Experiment}
collaboration)\\
$^1$Institute of Astronomy, University of Cambridge,
Madingley Road, Cambridge CB3 0HA, UK\\
$^2$European Organization for Nuclear Research CERN,
CH-1211 Gen\`eve 23, Switzerland\\
$^3$Institut f\"ur Theoretische Physik, Universit\"at Z\"urich,
Winterthurerstrasse 190, CH-8057 Z\"urich, Switzerland\\
$^4$Astronomy Unit, School of Mathematical Sciences, Queen Mary,
University of London, Mile End Road, London E1 4NS, UK\\
$^5$Universit\'e Bretagne-Sud, Campus de Tohannic, BP 573,
F-56017 Vannes Cedex, France\\
$^6$Laboratoire de Physique Corpusculaire et Cosmologie, Coll\`ege de
France, 11 Place Marcelin Berthelot, F-75231 Paris, France\\
$^7$Department of Astronomy, Ohio State University,
140 West 18th Avenue, Columbus, OH 43210, USA\\
$^8$Astrophysics Research Institute, Liverpool John Moores University,
12 Quays House, Egerton Wharf, Birkenhead CH41 1LD, UK}
\date{Accepted 2004 March 3;
Revised version submitted 2004 February 22 (v3), \& February 9 (v2);
Original version submitted 2004 January 19 (v1)}
\pagerange{\pageref{start}--\pageref{finish}}\pubyear{2004}

\begin{document}
\label{start}
\maketitle
\begin{abstract}
For the purposes of identifying microlensing events, the POINT-AGAPE
collaboration has been monitoring the Andromeda galaxy (M31) for three
seasons (1999-2001) with the Wide Field Camera on the Isaac Newton
Telescope. In each season, data are taken for one hour per night for
roughly sixty nights during the six months that M31 is visible. The
two 33\arcmin$\times$33\arcmin\ fields of view straddle the central
bulge, northwards and southwards. We have calculated the locations,
periods and brightness of 35414 variable stars in M31 as a by-product
of the microlensing search. The variables are classified according to
their period and brightness. Rough correspondences with classical
types of variable star (such as population I and II Cepheids, Miras
and semi-regular long-period variables) are established. The spatial
distribution of population I Cepheids is clearly associated with the
spiral arms, while the central concentration of the Miras and
long-period variables varies noticeably, the brighter and the shorter
period Miras being much more centrally concentrated.

A crucial role in the microlensing experiment is played by the
asymmetry signal -- the excess of events expected in the southern or
more distant fields as measured against those in the northern or
nearer fields. It was initially assumed that the variable star
populations in M31 would be symmetric with respect to the major axis,
and thus variable stars would not be a serious contaminant for
measuring the microlensing asymmetry signal. We demonstrate that this
assumption is not correct. All the variable star distributions are
asymmetric primarily because of the effects of differential extinction
associated with the dust lanes. The size and direction of the
asymmetry of the variable stars is measured as a function of period
and brightness. The implications of this discovery for the successful
completion of the microlensing experiments towards M31 are discussed.
\end{abstract}

\section{Introduction}

Hodge (1992) reports in his book on the Andromeda Galaxy that
\emph{``the variable stars of M31 have been the subject of a number of
rather limited studies. The very brightest are fairly well understood,
but only a few areas have been searched deeply. There is still an
immense amount of information about this component of the Andromeda
Galaxy that remains untouched.''}

\begin{figure}
\includegraphics[width=\hsize]{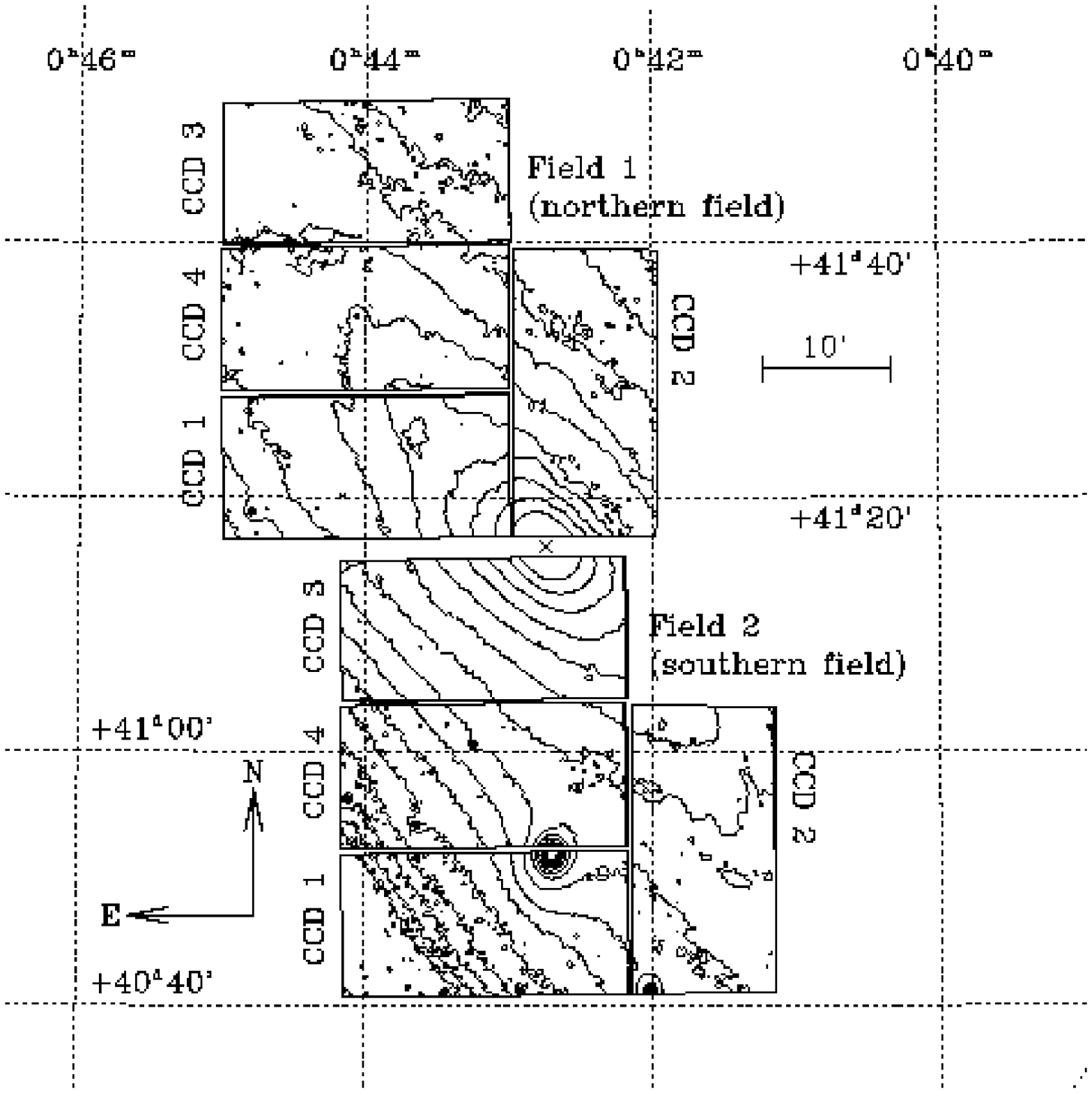}
\caption{The INT WFC observation fields of M31. Also drawn is the $r$
band surface brightness contour of M31 in the fields. The centre of
M31 is at $\alpha=0\h42\m44\fs31$, $\delta=+41\degr16\arcmin09\farcs4$
(J2000.0) and is marked by a cross.}
\label{fig:field}
\end{figure}

There is early work on Cepheids in M31 by Baade \& Swope (1963, 1965).
They found and studied $\sim$400 Cepheids, primarily for the purpose
of estimating the distance modulus for M31. A modern day successor to
this pioneering work is the DIRECT project (Stanek et al.\ 1999;
Mochejska et al.\ 2000; Bonanos et al.\ 2003). They have been
searching M31 (and M33) for detached eclipsing binaries and Cepheids
to refine the extragalactic distance ladder. They analysed five
11\arcmin$\times$11\arcmin\ fields in M31 and found a total of 410
variables of which roughly half were Cepheids. There is also a
substantial body of work on classical novae. Positions, magnitudes and
light curves already exist for $\sim$300 classical novae in M31
(Rosino 1973; Rosino et al.\ 1989; Ciardullo, Ford \& Jacoby 1983).
Despite all this, the underlying truth of the quotation from Hodge's
book is evident. The Andromeda galaxy is sufficiently close that the
variable star populations can be studied in enormous detail -- and
with much greater ease than those in our own Galaxy, for which an
all-sky survey would be needed.

Interest in the variable stars of M31 was re-kindled by influential
papers by Crotts (1992) and Baillon et al.\ (1993) on the possible
microlensing signal of the halo. The high inclination of the M31 disk
($i\approx 77\degr$) means that lines of sight to sources in the far
disk are longer and pass closer to the denser central parts of the
halo than lines of sight to sources in the near disk. In other words,
there is a strong gradient in the microlensing signal across the disk,
if the events are caused by objects in a spheroidal halo. Early on, it
was argued that the stellar populations of M31 are well-mixed and so
no variable star population would show this behaviour (Crotts 1992).
Thus, it seemed that the strategy in the experiment was
straightforward; first, microlensing candidates should be identified
on the basis of achromaticity and goodness of fit to a Paczy\'nski
curve, and second the gradient of detected candidates should be
measured and confronted with theoretical models. The asymmetry in
microlensing rate between the near and the far disk therefore seemed
an unambiguous way of diagnosing events caused by lenses in the halo,
as opposed to stellar lenses, and thus resolving the ambiguity in the
results of the experiments monitoring the Large Magellanic Clouds.
Accordingly, a number of groups began intensive monitoring of M31 with
the aim of detecting microlensing events, including the POINT-AGAPE
(Auri\`ere et al.\ 2001; Paulin-Henriksson et al.\ 2002, 2003; Calchi
Novati et al.\ 2003; An et al.\ 2004), WeCAPP (Riffeser et al.\ 2001,
2003) and MEGA (de Jong et al.\ 2004) collaborations.

The first hint that this was not the case came when Burgos \&
Wald-Doghramadjian (2002), working with data taken by the POINT-AGAPE
collaboration, showed that the numbers of variable stars in
symmetrically positioned fields in the near and far disk were not
equal. The implication of this finding made the microlensing
experiments much more difficult, as the asymmetry in the variable
stars needs to be studied and measured as a function of brightness and
period. Accordingly, the analysis of the variable star database in the
microlensing experiments towards M31 is interesting from the
standpoint of the light it throws on stellar populations in M31 and is
crucial for a successful conclusion to the microlensing experiments.

The purpose of this paper is to present the variable star catalogue of
the POINT-AGAPE survey. This gives the position, period and brightness
of 35414 variable stars with high signal-to-noise lightcurves
uncovered during the three-year duration of our survey. The variable
star catalogue forms a subset of the grand total of 97280 lightcurves
that show variations. The paper is arranged as follows:
\S~\ref{sec:data} describes the acquisition of the data,
\S~\ref{sec:image} discusses the images of M31 and the detection of
resolved stars, \S~\ref{sec:catalogue} outlines the construction of
the lightcurves and the selection of the variable stars for the
catalogue, \S~\ref{sec:property} studies the properties of the
variable stars, whilst \S~\ref{sec:asymmetry} analyses the asymmetry
signal and discusses the consequences for the microlensing surveys.

\begin{table}
\caption{The date, exposure, and seeing for the $r$ band reference
images for the northern and southern fields.}
\label{table:reference}
\begin{tabular}{lcccccl}
\hline
Field & CCD & Image & Exposure & Seeing & Date \\
\hline
northern & 1 and 2 & r188970 & 700 s & 1\farcs6 & 14/08/1999\\ 
northern & 3 and 4 & r197399 & 360 s & 1\farcs4 & 05/10/1999\\
southern & 1 and 2 & r188986 & 700 s & 1\farcs4 & 14/08/1999\\
southern & 3 and 4 & r187566 & 600 s & 1\farcs5 & 08/08/1999\\
\hline
\end{tabular}
\end{table}

\section{Data Acquisition}
\label{sec:data}

Observations of M31 were obtained using the Wide Field Camera (WFC)
mounted on the 2.5~m Isaac Newton Telescope (INT) located at La Palma,
the Canary Islands, Spain, over three observing seasons, from 1999
August to 2002 January. Three broad-band filters, $g$, $r$ and $i$,
were employed. The combination of the filters and the response of the
thinned EEV 4k$\times$2k CCDs of the WFC produce overall pass-bands
similar, but not identical, to the Sloan Digital Sky Survey $g'$, $r'$
and $i'$ bands. The $r$ band images were taken throughout the 3-year
campaign. Initially, a mix of both $g$ and $i$ band images were also
obtained, but the $g$ band monitoring was discontinued at the end of
the first season. Weather and technical conditions allowing,
observations in the second and third seasons consisted of both $r$ and
$i$ band exposures in each of the two M31 fields.

\begin{figure*}
\vspace{0.27\hsize}{\tt 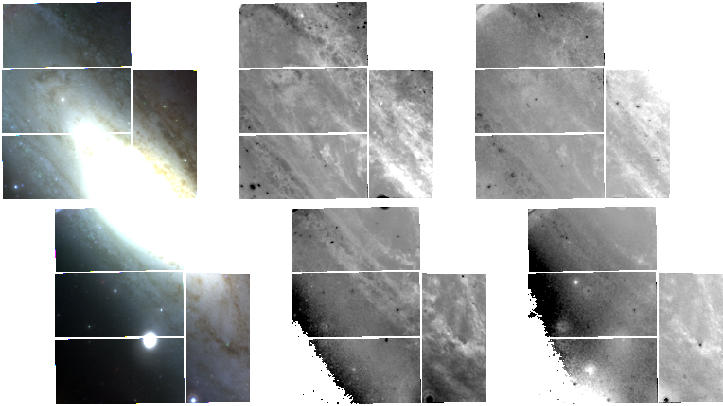}\vspace{0.27\hsize}
\caption{The INT WFC $gri$ image (left) and $g-r$ (centre) and $r-i$
(right) colour map of M31. One can observe very prominent dust
structure in the northwestern part of M31. The (infra)red sensitivity
or the background of CCD 2 may be slightly higher than other chips.}
\label{fig:image}
\end{figure*}

The data analysed here consist of the $r$ and $i$ band exposures
obtained over all three seasons. The location of the two fields, to
the north and south of the nucleus of M31 can be seen in
Figure~\ref{fig:field} (see also Fig.~5 of Paulin-Henriksson et al.\
2003). Typically, pairs of exposures were obtained in each band in
each field in the sequence $r$ field 1, $i$ field 1, $r$ field 2 and
$i$ field 2. Exposure times, normally 320 s in both $r$ and $i$ bands,
were chosen to maximise the signal-to-noise ratio of each exposure,
while ensuring that two exposures in each band could be obtained
within the 3600 s allocated to the monitoring of M31 on each
night. Advantages of acquiring two exposures per band per field
included the ability to identify cosmic rays and to avoid saturation
in exposures obtained in the $i$ band during full moon.

A standard observing script was invoked to perform the observations
and the data are generally of high quality. The on-chip seeing
achieved falls predominantly in the range 1\farcs3-2\farcs0, which
is well-sampled by the 0\farcs33 pixels of the WFC. Observations on a
small number of nights were affected by technical problems that
resulted in frames with highly elongated images. Such frames were
excluded from the analysis described here.

Monitoring of M31 was possible only when the WFC was mounted on the
INT and when the telescope was scheduled for use by either the United
Kingdom or Netherlands time allocation committees. We were fortunate
to also obtain occasional monitoring observations from a few
Spanish-scheduled observers. The monitoring schedule was further
enhanced through observations obtained by INT staff and others during
nights scheduled for engineering activities and service-observing.

The cooperation of astronomers and staff at the INT resulted in a
total of 71 $g$, 316 $r$, and 254 $i$ band for the northern field and
67 $g$, 293 $r$, and 232 $i$ band for the southern field monitoring
observations that produced frames suitable for analysis. However, the
vagaries of the weather, combined with the rigid schedule for mounting
of the WFC, means that the monitoring time-series consists
predominantly of relatively well-sampled periods, typically of
duration 7-14 days, interspersed with periods, typically of duration
7-21 days, in which there are very few data. A summary of the
observations taken is publicly available over the World Wide
Web.\footnote{\tt\scriptsize
http://cdfinfo.in2p3.fr/Experiences/Agape/point-agape\_log.html}

All frames were processed using the pipeline processing software
(Irwin \& Lewis 2001) developed for the reduction of the data obtained
as part of the Wide Field Survey (WFS) undertaken with the WFC. The
pipeline includes procedures to de-bias, trim and flatfield the
frames. Additionally, protocols to perform non-linearity corrections,
flag bad pixels, and apply gain corrections are applied. Master
calibration frames (bias and flatfields, for $r$ and $i$ exposures,
and fringe frames, for $i$ exposures) necessary to reduce the
observations, were constructed from all the suitable calibration
frames obtained during the relevant 14-21 day period that the WFC was
mounted on the INT. On a few occasions, insufficient calibration
frames were obtained to construct an adequate master flatfield or
fringe frame. In these cases the best master calibration frames from
an adjacent period, when the WFC was mounted, were used.

The WFS pipeline proved entirely satisfactory with the exception of
the defringing correction necessary for the $i$ band frames. The WFS
defringing scheme was developed to cope with exposures whose
background is dominated by sky. In the case of our M31 exposures the
``background'' includes a significant contribution from M31 itself.
Furthermore, the relative contribution of M31 to the background varies
from CCD to CCD and between the exposures of the northern and southern
fields. The variable sky brightness over a lunation also complicates
the situation, as the contrast between M31 and the underlying
sky-background changes.

\begin{figure*}
\includegraphics[width=15cm]{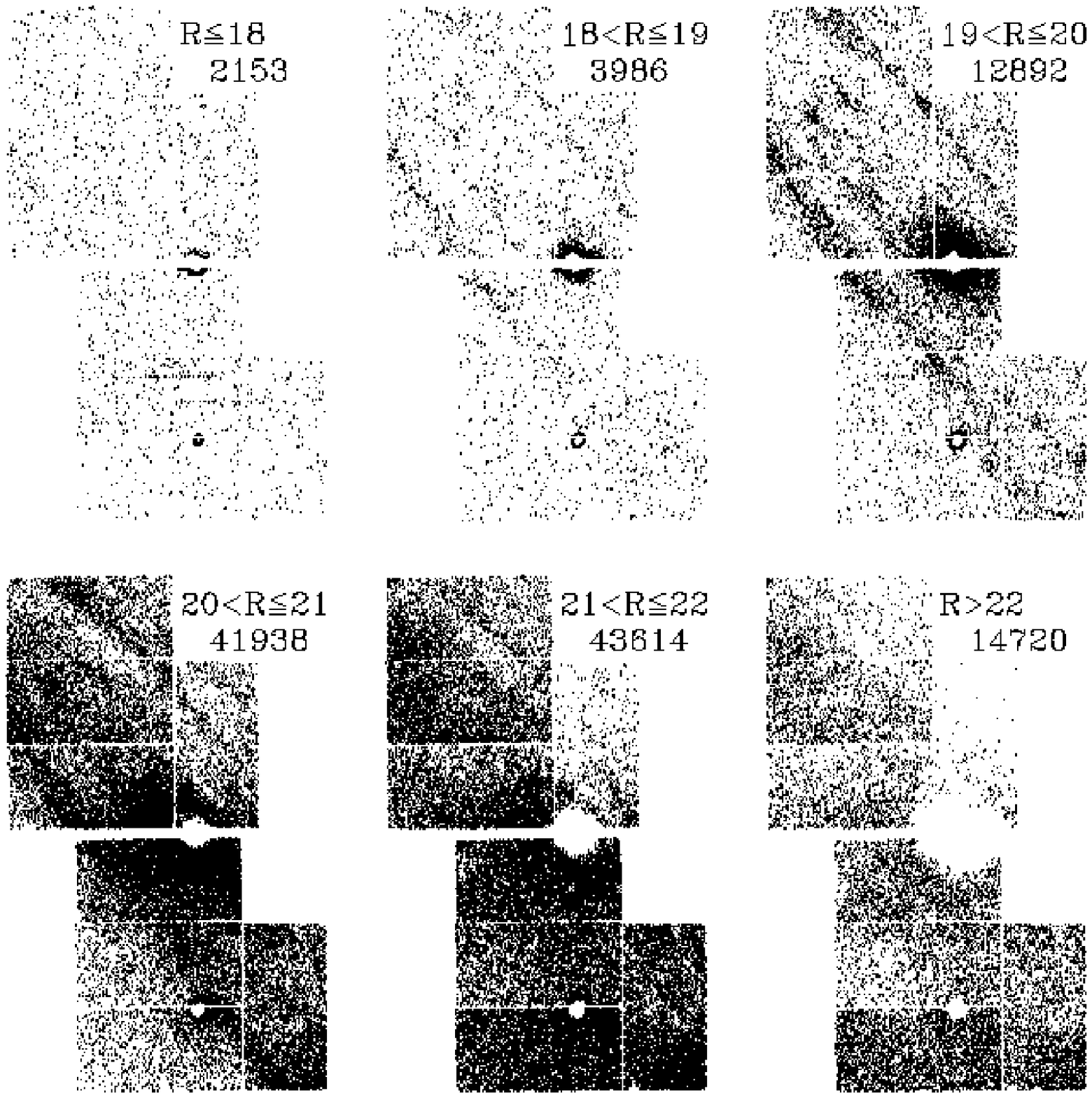}
\caption{Spatial positions of resolved stars with different
magnitudes. The stars with $R\le 18$ form an almost uniform screen
and so are predominantly foreground contaminants. There is also no
evidence of chip to chip variation for stars with $R\le 21$,
suggesting that the catalogue is complete to this limit.}
\label{fig:distr}
\end{figure*}

Investigation of the amplitude of the defringing correction made by
the WFS pipeline to remove the fringing from frames taken away from
M31 showed a well-defined correlation with the background surface
brightness. Small, $<10\%$, systematic offsets in the amplitude of the
scaling applied to the master fringe frames from CCD to CCD were also
evident. Comparison of the underlying sky-brightness in frames (away
from M31) taken immediately prior or subsequent to the M31 monitoring
showed that an accurate estimate of the underlying sky-brightness in
the M31 frames could be obtained using the far south-eastern corner of
CCD 1 in the southern field. Thus, the amplitude of the defringing
correction applied to both the northern and southern M31 frames was
derived using the estimate of the sky-brightness from the
south-eastern corner of CCD 1 in the southern field, allowing for the
well-determined small systematic offsets from CCD to CCD.

\vfill
\section{The surface brightness and the resolved stars of M31}
\label{sec:image}

Figure~\ref{fig:image} shows a $gri$ (which is close to the standard
$VRI$) image of M31, together with the $g-r$ and $r-i$ (close to $V-R$
and $R-I$ respectively) colour maps. These are constructed by median
filtering the images with a 49 pixel (16\farcs3) square box and
re-sampling with a 10\arcsec\ grid. The filtering is done to remove
graininess in the images, while the re-sampling is done to enable a
fair comparison with maps of the variable star distribution which will
be presented later. The colour maps clearly delineate the dust lanes.
It is evident that the dust distribution is asymmetric with the
north-western parts of M31 being dustier. This is already worrisome,
as it may muddle any intrinsic microlensing asymmetry.

\begin{figure*}
\includegraphics[width=15cm]{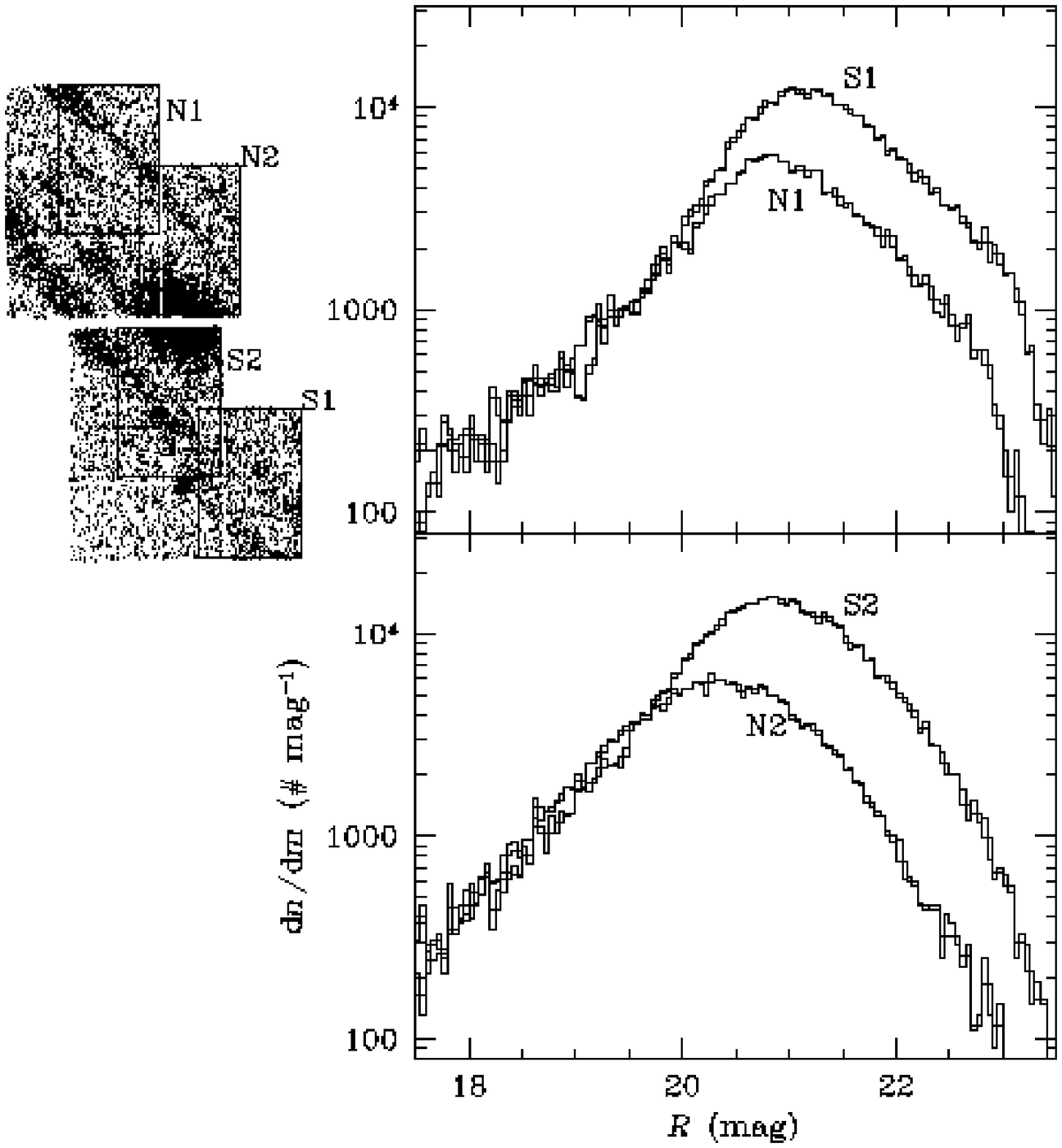}
\caption{The observed uncorrected luminosity function in pairs of
fields (namely N1, S1 and N2, S2) symmetrically positioned with
respect to the centre of M31. All four fields are
15\arcmin$\times$22\arcmin\ in size.}
\label{fig:reshist}
\end{figure*}

We also construct a catalogue of 119303 resolved stars -- namely, all
stars detected by the task {\it daofind} in IRAF at 5$\sigma$ above
the background noise in the $r$ band reference image list in
Table~\ref{table:reference}. For the detected stars, the $r$
magnitudes are converted into standard Cousins $R$ magnitudes.
Figure~\ref{fig:distr} shows the break-down of the 119303 resolved
stars into magnitude bins. Stars with $R\le 18$ are predominantly
foreground objects, as evidenced by their homogeneity. For $18\la R\la
20$, we can see a mixture of M31 stars -- which are predominantly
young, massive supergiants -- together with some contamination from
the foreground and from unresolved globular cluster cores and blended
stars. The supergiant population is concentrated in star-forming
regions associated with the spiral arms, which are clearly seen in
Figure~\ref{fig:distr} and coincide with the blue region in
Figure~\ref{fig:image}.

For $R\ga 20$, the asymptotic giant branch (AGB) stars start to make
their appearance, while the red giant branch (RGB) stars are somewhat
fainter and do not appear at least up to $R\simeq 21$. With the
absolute magnitude of the tip of the RGB being $M_I\simeq -4.0$
(Bellazzini, Ferraro \& Pancino 2001), and the distance modulus of M31
being $(m-M)_{\rm M31}\simeq 24.5$ (Holland 1998; Stanek \& Garnavich
1998), the brightest RGB stars are around $R\approx 21.5$, assuming
the colour $R-I\ga 1.0$. The brightest AGB stars are around a
magnitude or so brighter. Again, on comparing Figure~\ref{fig:distr}
with Figure~\ref{fig:image}, we see that there is a clear depression
in the distribution of AGB stars (with $20<R\le 21$) coincident with
the prominent dust lane in the north-west of the galaxy.

The limiting magnitude of field 1, CCD 2 seems to be somewhat
shallower than the other CCDs. We can also see some evidence of
variation of the limiting magnitude with surface brightness near the
central bulge and near M32. This therefore raises the worry that
variations in the limiting magnitude from CCD to CCD, caused by the
choice of reference images with different background and exposure, may
be producing artificial gradients. To investigate this, we plot in
Figure~\ref{fig:reshist} the distribution of resolved stars in
different magnitude bins, or the observed luminosity function. Here,
the two pairs of fields (N1, S1 and N2, S2) are chosen to be symmetric
with respect to the centre of M31, as illustrated in the side-panel of
the figure. For each field, two nearly-superposed histograms are
shown, one for a bin-size of 0.1 mag, one for a bin-size of 0.05 mag.
The excellent correspondence between the histograms shows that the
bin-size is not influencing the results.

\begin{figure*}
\vspace{0.27\hsize}{\tt 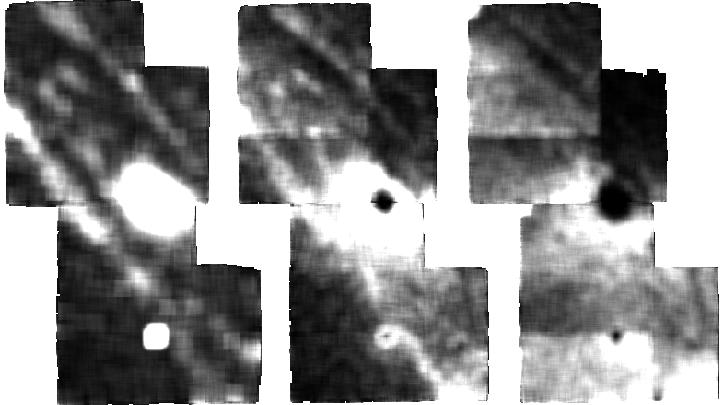}\vspace{0.27\hsize}
\caption{Grey-scale surface density maps of the resolved stars;
$R\le20$ (left), $20<R\le21$ (centre), $R>21$.
\hspace{\hsize}}
\label{fig:resmap}
\end{figure*}

The luminosity functions in all four fields turn over at $R\sim 20.5$,
and are in good agreement for $R<20$. The turn-over in the luminosity
function is almost certainly due to incompleteness. However, we are
interested primarily in comparisons between the northern and southern
fields, and so provided we believe that roughly the same fraction of
stars is missing, we can go deeper than the turn-over.

For $R\ga 20$, the luminosity function in the southern fields (S1 and
S2) is higher than in their northern counterparts (N1 and N2). For
higher surface brightness, the detection limit should be shallower.
However, the reverse is true here, as the southern fields are brighter
-- not fainter -- than the northern ones. This is circumstantial
evidence that the relatively low normalisation of the northern field
luminosity function is a real effect, rather than an artefact of
differing efficiency. Note that this rules out the lower sensitivity
of CCD 2 as a cause of any asymmetry, as we would then expect more
stars in N1 than S1, which is not seen. However, it is still possible
that CCD 2 in field 1 alone is the cause of the problem because of the
underlying surface brightness structure in that field. Part of the
low normalisation is probably caused by the prominent dust lane and
enhanced extinction in the northern field. Additionally, the slope of
the falling parts of each pair of histograms at least for $R\la 22$
in Figure~\ref{fig:reshist} is the same. If the underlying true
luminosity function has the same slope, then this can be explained
very naturally by assuming that the same fraction of stars is missing
at each magnitude, irrespective of whether the field is north or south.

For the purposes of our analysis, it is helpful to convert the spatial
distribution shown in Figure~\ref{fig:distr} into smooth maps of the
surface number density. This is achieved by counting the number of
resolved stars within a suitable window function. In
Figure~\ref{fig:resmap}, this is done for three different magnitude
bins. In the left panel corresponding to resolved stars with $R\le
20$, the window function is a square box of size 3\arcmin. For the
other two panels, the square box has size 2\arcmin. The re-sampling is
done in the same way as described for Figure~\ref{fig:image}. Again,
for $R<20$, we see the prominent ring (possibly a spiral arm)
associated with the supergiant population. Some of the dust structures
visible in the colour maps of Figure~\ref{fig:image} can be traced in
the surface density maps of the fainter stars. There are also
artefacts -- such as the deficiency in resolved stars with $R>20.5$ in
field 1, CCD 2 -- that are manifest in the Figure.

To conclude, there is clear asymmetry in the resolved star
distribution, and some of this is certainly due to bias during the
data reduction. However, the similar patterns caused by dust lanes
clearly visible in the surface brightness map, the colour maps and the
resolved star maps argue that there is a real asymmetry caused by
variable extinction. This hypothesis could be confirmed by examination
of the colour-magnitude diagrams in the northern and southern fields,
which should show differing positions for the RGB stars.

\section{Construction of the Variable Object Catalogue}
\label{sec:catalogue}

The M31 fields are composed of largely unresolved stars, and so the
effects of seeing are significant. In order to build lightcurves, we
need to develop an algorithm for which the same fraction of flux falls
within the window function, irrespective of seeing. The superpixel
method provides a linear transformation between the flux measurements
with different seeing (Melchior et al.\ 1999; Ansari et al.\ 1999; Le
Du 2000). The basic idea is to calibrate the frames so that they have
the same large-scale average surface brightness and then to adjust any
spatial flux variations to the same seeing.

\begin{figure}
\includegraphics[width=\hsize]{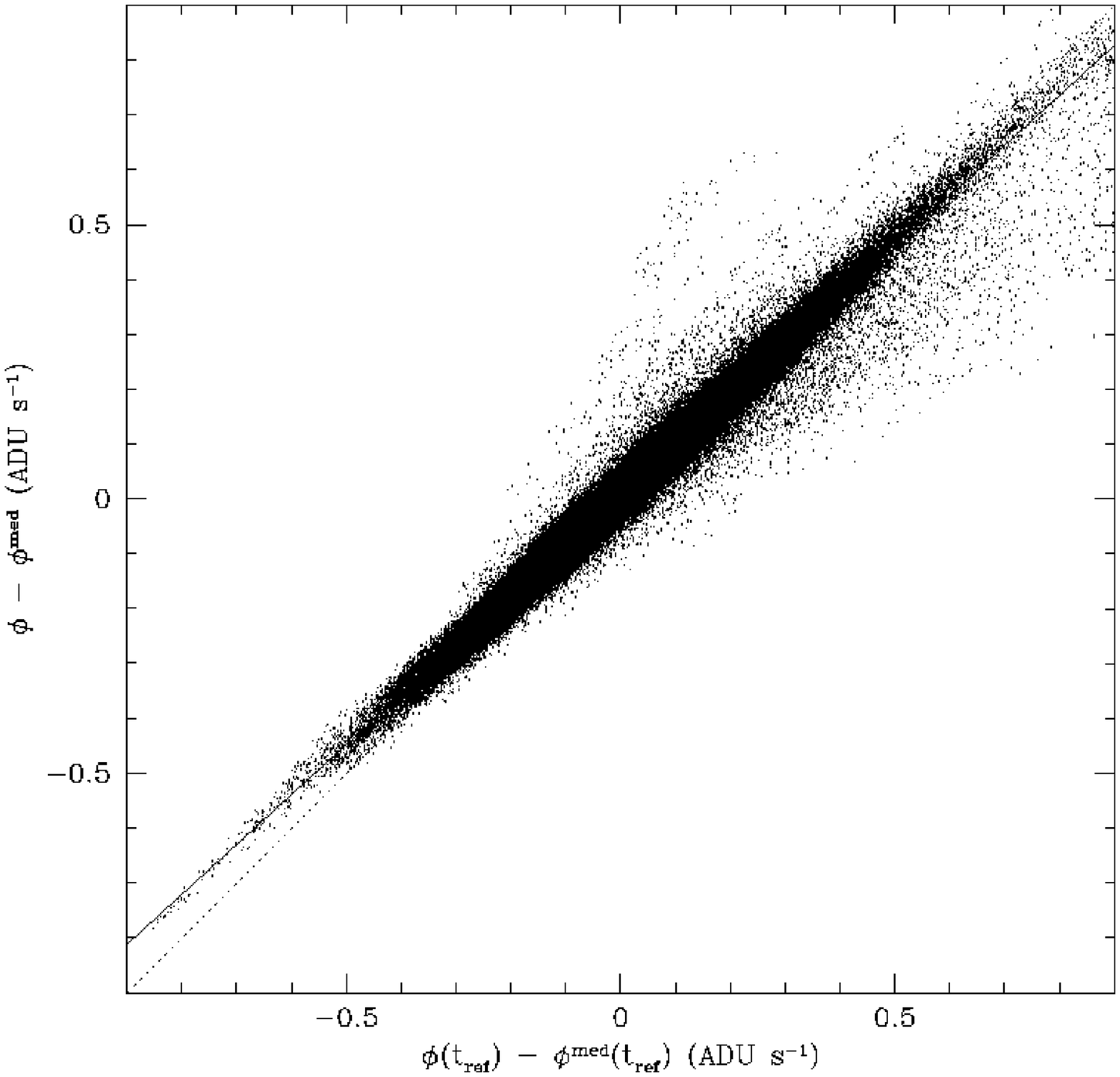}
\caption{The linear correlation between flux differences on the
reference image and on an image with good seeing respectively. Each
dot in the ``cigar'' corresponds to a superpixel. The solid line is
the best fit linear regression line, whilst the dotted line
illustrates the identity relation.}
\label{fig:cigar}
\end{figure}

The observed flux $\phi$ within the window function at a given
position is a convolution of the intrinsic surface brightness $\Phi$
with the point-spread function (PSF) $P$ and the window function $W$.
Assuming that the PSF is flux-conserving, the large scale average flux
$\bar\phi=\bar\Phi\otimes W$ is independent of the PSF. Here,
$\bar\Phi$ is the large scale average of the intrinsic surface
brightness. Furthermore, the spatial fluctuation of the flux is also
related to the intrinsic spatial surface brightness fluctuation via
the same convolution, viz.\
\begin{equation}
\phi-\bar\phi=(\Phi-\bar\Phi)\otimes P\otimes W.
\label{eq:primitive}
\end{equation}
Taking Fourier transforms, the convolutions become products, and so we
have
\begin{equation}
\cf\{\delta\phi\}=\cf\{\delta\Phi\}\cf\{P\}\cf\{W\},
\label{eq:fourier}
\end{equation}
where $\cf$ denotes the Fourier transform. The effects of the shape of
the PSF near the centre are unimportant, if the characteristic scale
of the window function is chosen to be large enough. On the other
hand, for any reasonable PSF, the wings behave in a similar manner to
lowest order. Then, taking the ratio of equations~(\ref{eq:fourier})
for two epochs implies that the Fourier transforms of the flux
variations are linearly related. Consequently, the flux differences at
different epochs are also linearly related. The superpixel method
provides an empirical way of calculating the coefficients of this
linear transformation. Note that this linearity is only true in
lowest order, and so the method fails for large seeing variations or
near very bright stars.

The original pixel size (0\farcs33) over-samples the seeing disk
($\mbox{FWHM}\ga 1\arcsec$), and thus the relative amplitude of the
flux variations due to seeing variation is quite large even for
moderate variation of seeing. This leads us to associate a superpixel
flux $\phi$ with each pixel. This is derived by summing fluxes from
several adjacent pixels that fall within a pre-defined ``superpixel''
window (in other words, $W$ is a top-hat function defined over a
square box). The size of this superpixel is chosen to ensure that it
under-samples the seeing disk.

\begin{figure}
\includegraphics[width=\hsize]{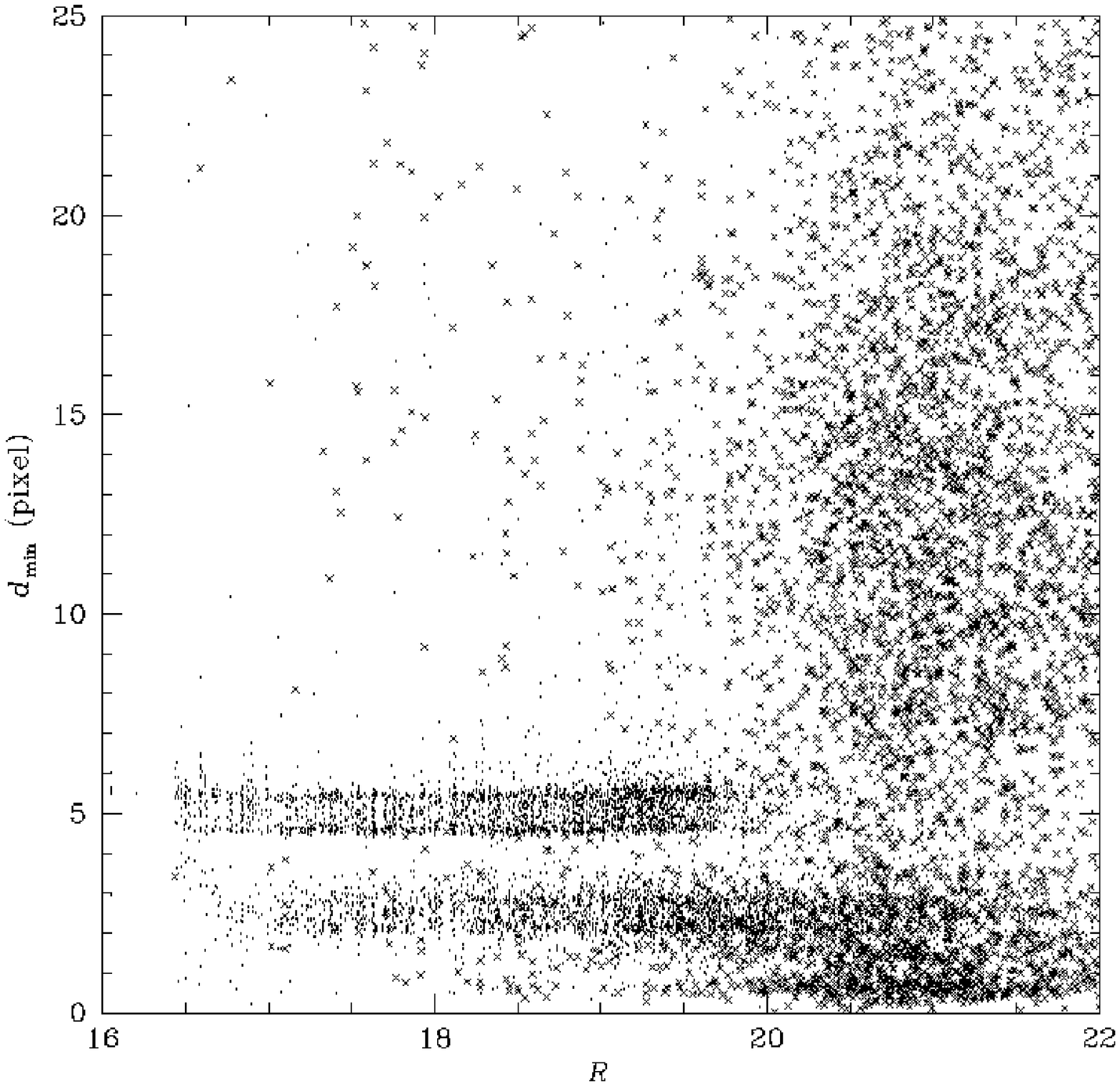}
\caption{Each point represents a variable lightcurve in CCD 4 of the
northern field. The $R$ magnitude of the nearest resolved star is
plotted against the distance to that resolved star. In other words, a
point at coordinates (18, 2) means that there is a variable star
lightcurve that lies 2 pixels away from a resolved star with $R=18$.}
\label{fig:fakes}
\end{figure}

Next, the data are smoothed by a median filter with a window size that
is sufficiently large so as not to be affected by small-scale
fluctuations. This gives the median superpixel flux $\phi^{\rm med}$
at each point. This median flux is subtracted from the superpixel
flux.\footnote{In our application, the median flux $\phi^{\rm med}$ is
used instead of the mean $\bar\phi$ in eq.~(\ref{eq:primitive}).} The
quantity $\phi-\phi^{\rm med}$ represents a variation in flux with
respect to the same underlying median surface brightness. It changes
with the seeing. Empirically, we find that the flux differences at
different epochs are linearly related to each other, as shown in
Figure~\ref{fig:cigar}. By recalibrating the flux differences, we
obtain superpixel lightcurves which represent the intrinsic variation
at fixed seeing. In practical implementation, we find the slope
$\alpha$ and the intercept $\beta$ of the regression line on the cigar
shape depicted in Figure~\ref{fig:cigar}, namely
\begin{equation}
\phi(t_i)-\phi^{\rm med}(t_i)=\alpha(t_i)
\left[\phi(t_{\rm ref})-\phi^{\rm med}(t_{\rm ref})\right]+\beta(t_i),
\end{equation}
and the subsequent recalibration follows as
\begin{equation}
\phi'(t_i)=
\frac{\left[\phi(t_i)-\phi^{\rm med}(t_i)\right]-\beta(t_i)}{\alpha(t_i)}
+\phi^{\rm med}(t_{\rm ref}).
\label{eq:seecorr}
\end{equation}

If the background subtraction is sufficiently accurate and if the
fluxes represent the intrinsic, physical surface brightnesses, then
the median flux must be constant, $\phi^{\rm med}(t_i)=\phi^{\rm med}
(t_{\rm ref})$. In this case, the linear relation will become a
proportionality ($\beta=0$), while the coefficient $\alpha$ will be
solely a function of the seeing difference (in particular, if the
images at $t_i$ and $t_{\rm ref}$ are taken in the same seeing, then
$\alpha$ will be unity).\footnote{Even if the photometric calibration
is imperfect, then the seeing correction~(\ref{eq:seecorr}) will still
give the correct results but with non-vanishing $\beta$.} For best
results, the reference image is chosen to be one taken in the median
seeing. We note that the linear relation shown in
Figure~\ref{fig:cigar} is not a perfect one-to-one correspondence, but
is rather an (albeit tight) statistical correlation with a scatter. In
other words, this procedure of stabilising the seeing variation
introduces an additional uncertainty. Therefore, the error bar on the
$i$-th flux measurement, corrected for seeing, is
\begin{equation}
\sigma_i^2=\frac{1}{\alpha^2}{\sigma_\gamma}_i^2+\sigma_{\rm see}^2,
\end{equation}
where $\sigma_\gamma$ is the photon noise in the original image and
$\sigma_{\rm see}$ is the additional uncertainty caused by the
recalibration. This is empirically estimated from the width of the
cigar shape in Figure~\ref{fig:cigar} using the method of maximum
likelihood.

\begin{figure}
\includegraphics[width=\hsize]{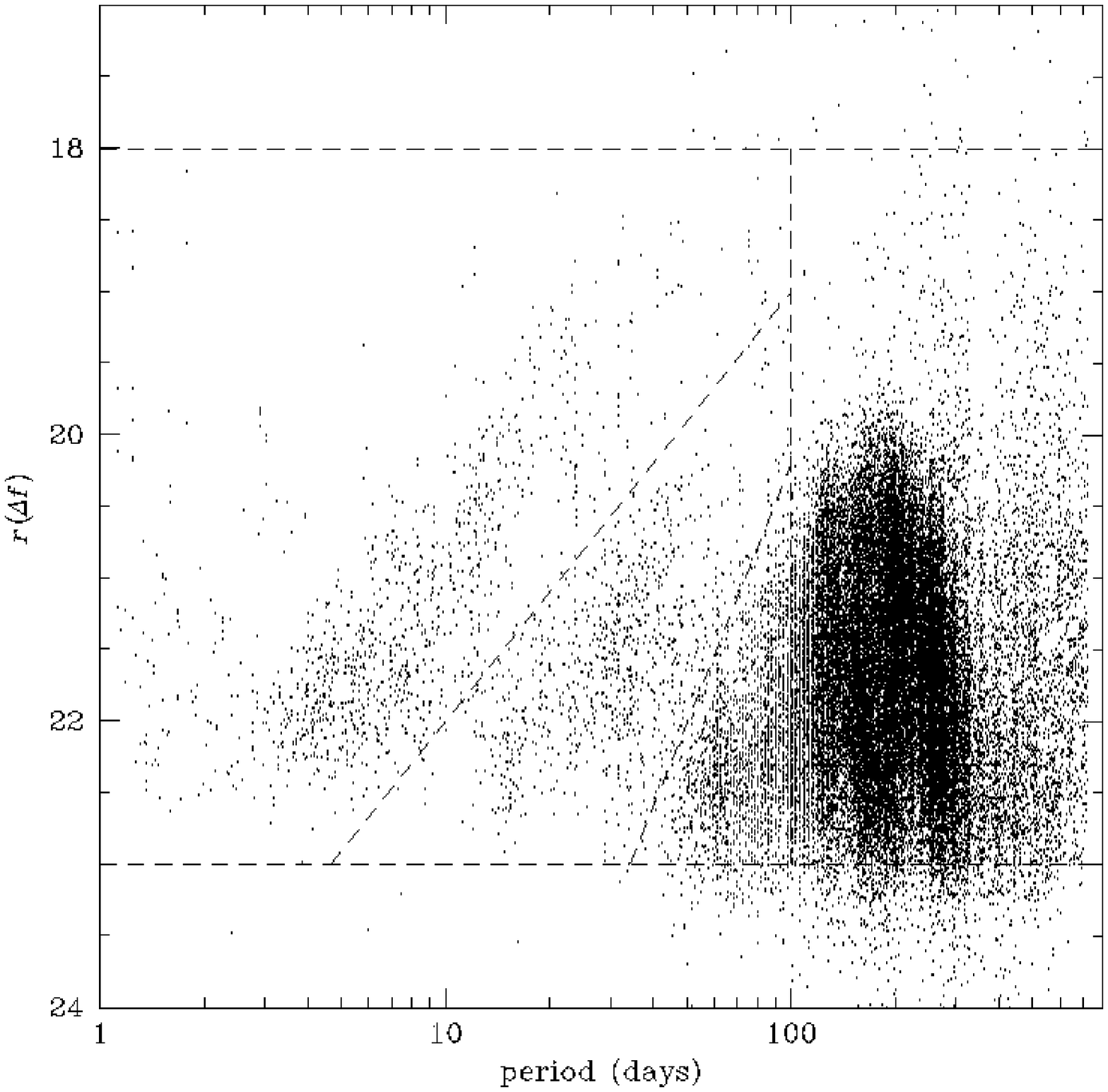}
\caption{The locations of the $\sim$40,000 variable stars in the
period-$r(\Delta f)$ plane. The period computed via Lomb's periodogram
is discrete and so the positions are shifted by small random values to
enhance visibility. Notice the two straight-line features in the left
of the plot, which are suggestive of period-luminosity relationships.
Also shown with dashed lines are the boundaries for the group classification
given in the text.}
\label{fig:peramp}
\end{figure}

In the implementation of the method here, the median flux is computed
over a 41$\times$41 pixel field. The method is not too sensitive to
the size of the field, provided it is big enough that even bright
resolved stars disappear from the median image. There is an optimum
size to the superpixel. If the superpixel is too small, the algorithm
fails because the seeing disk is over-sampled. However, as the
superpixel gets larger, the photon noise increases and the sensitivity
drops. We choose the superpixels to be 7$\times$7 pixels in size.
This is because typical poor seeing at the INT site is $\sim$2\arcsec\
so that a star with a Gaussian PSF is contained up to a 1$\sigma$
distance in the 7$\times$7 superpixel centred on it. The reference
images used for the $r$ band data are the same as the image on which
the detection of resolved stars has been performed
(Table~\ref{table:reference}).

The WFC has 2k$\times$4k pixels in each chip. There are 4 chips and 2
fields, giving $\sim 6.4\times 10^7$ superpixel lightcurves for each
passband. The selection procedure is carried out on the $r$ band
data alone. For each lightcurve, the first step is to determine the
baseline flux $\mu$. This is calculated iteratively by computing the
median, deleting all points 3$\sigma$ above (but not below) the
baseline and repeating until convergence. This is tailored for
microlensing events, which are positive excursions above the baseline,
but fails for eclipsing binaries or transits. Any lightcurve that has
at least 3 consecutive points at least 3$\sigma$ above the baseline is
selected as a fluctuation. For each variable object, there is a
cluster of associated superpixel lightcurves. The likelihood $L$ of
variation for every superpixel position is calculated by
\begin{equation}
\ln L=\sum_{i\in{\rm fluct}}\ln P(\phi_i),
\end{equation}
with
\begin{equation}
P(\phi_i)=\frac{1}{(2\upi)^{1/2}\sigma_i}
\int_{\phi_i}^\infty\!{\rm d}\phi\ 
\exp\left[-\frac{1}{2}\left(\frac{\phi-\mu}{\sigma_i}\right)^2\right].
\end{equation}
For each superpixel, a likelihood map is constructed over the field
and the optimum position of each variable source is extracted from it
using SExtractor (Bertin \& Arnouts 1996).

\begin{figure}
\includegraphics[width=\hsize]{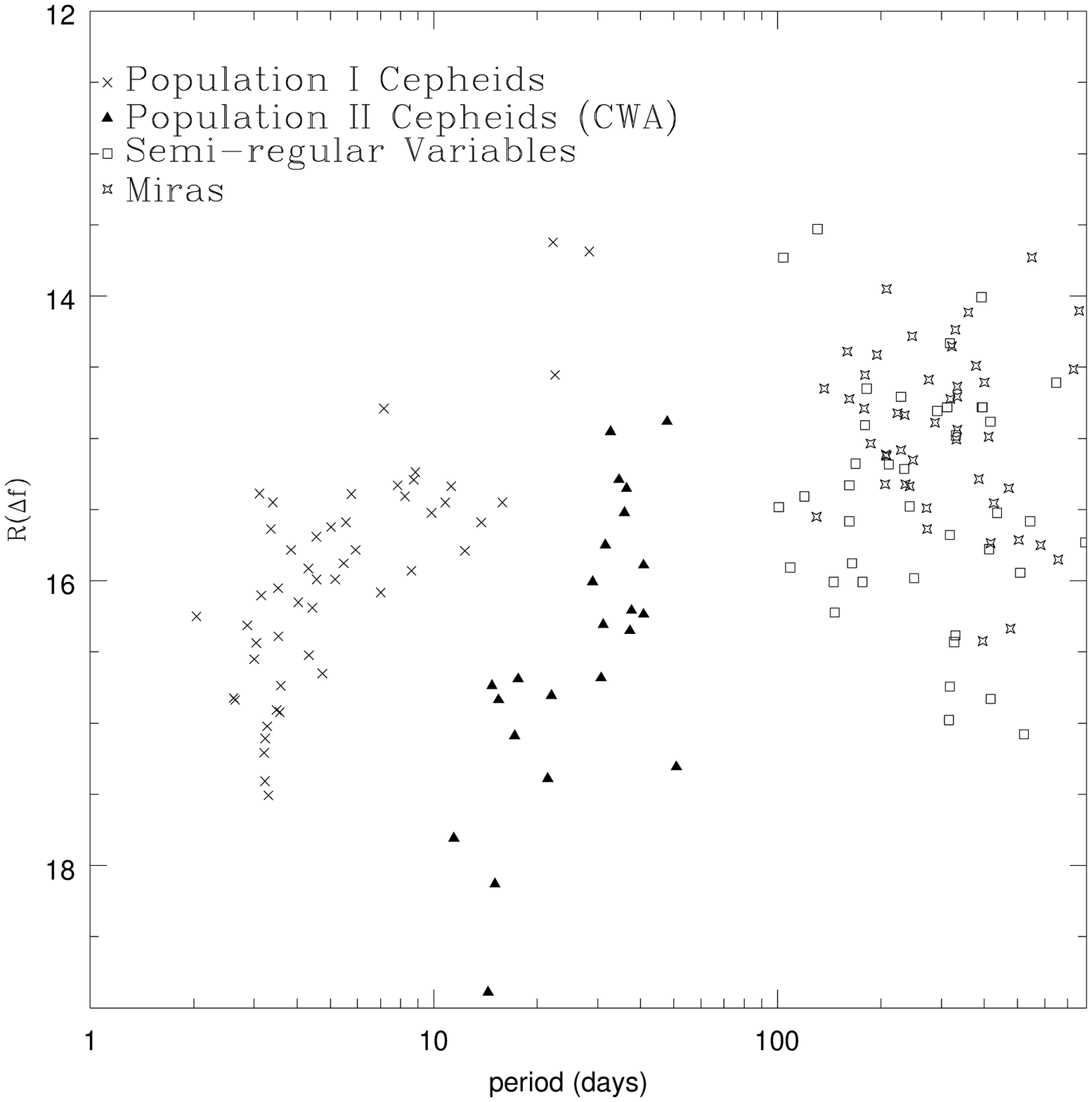}
\caption{The locations of some common types of variable stars in the
Large Magellanic Cloud plotted in the period-$R(\Delta f)$ plane. The
data are drawn from the General Catalogue of Variable Stars.}
\label{fig:varstar}
\end{figure}

%Any source found to lie at the position of a bright resolved
%star or known CCD defect is discarded.

We select a total of 97280 lightcurves by the procedure outlined
above. However, many lightcurves in this list are spurious; most of
the artefacts are caused by a bright resolved star lying near the edge
of a given superpixel, which induces a non-linear response of the
adjacent superpixel flux level depending on the variation of seeing.
Since the seeing varies without any significant coherence on a nearly
daily basis, the resulting lightcurves appear almost like white noise
although the amplitude can be many times larger than the typical
photometric uncertainty of the superpixel flux. Such artefacts can be
recognised easily by eye. It is however clearly preferable to devise
an objective criterion. Referring to Figure~\ref{fig:fakes}, we see
that the there is an excess of variables around 2.5 and 5 pixel
distance from resolved stars, which corresponds to the concentration
of artefacts. These can be removed by construction of a suitable mask
consisting of annuli around resolved stars and known CCD defects. This
is done for the selection of microlensing candidates in a later
publication. In this paper, however, we use a different (but related)
criterion, which is explained in the next section.

\vfill
\section{The Properties of the Variable Stars}
\label{sec:property}

We characterise each variable lightcurve by two numbers, its period
$P$ and the ``pseudo-magnitude'' $r(\Delta f)$. The period is computed
using Lomb's periodogram (e.g., Press et al.\ 1992), which is
appropriate for unevenly sampled datastreams. The algorithm works by
calculating the power spectrum over a discrete frequency interval. The
frequency at maxmimum power is converted into the period for the
lightcurve.  Therefore, the absolute uncertainty in the period scales
quadratically with the period itself and so the computed value may not
be precise when $P\ga 150\ \mbox{days}$. In principle, there are other
methods of computing a more precise period, but Lomb's periodogram
suffices for the purpose of classification.

As a measure of the brightness of each variable, we define the
``pseudo-magnitude''. Because the fraction of light contributed by the
variable object to the total superpixel flux at a given epoch is not
known, it is not possible to find the real magnitude of the
variable. Under these circumstances, we first measure the quantity
$\Delta f$, which is the flux variation between minimum and maximum in
ADU s$^{-1}$. Then, we convert $\Delta f$ into magnitudes by $r(\Delta
f)=r_0-2.5\log_{10}(\Delta f)$, which we subsequently employ as a
proxy for the underlying brightness of the variable.  Here, $r_0$ is
the zeropoint, that is, the $r$ band magnitude of a star whose flux is
1 ADU s$^{-1}$ in the reference image.  Henceforth, we refer to this
quantity $r(\Delta f)$ -- or $i(\Delta f)$, $g(\Delta f)$ and so on
where appropriate -- as a pseudo-magnitude.  We note that it is not
the same as the variable amplitude in magnitudes (that is $\Delta
r\equiv r_{\min}-r_{\max}= 2.5\log_{10}[1+(\Delta f/f_{\min})]$ where
$r_{\min}$ and $f_{\min}$ are the magnitude and the flux at minimum,
and $r_{\max}$ is the magnitude at maximum). Rather, the
pseudo-magnitude is related to the magnitude of variable and the
variable amplitude via
\begin{eqnarray}
r(\Delta f)&=&r_{\max}-2.5\log_{10}[1-10^{-0.4\Delta r}]
\nonumber\\&=&r_{\min}-2.5\log_{10}[10^{0.4\Delta r}-1].
\end{eqnarray}
Therefore, as $\Delta r$ becomes larger, $r(\Delta f)$ approaches the
true magnitude at maximum $r_{\max}$ of the variable. [If $\Delta r=
0.5$, then $r(\Delta f)\approx r_{\max}+1.1$. On the other hand, if
$\Delta r=1.0$, then $r(\Delta f)\approx r_{\max}+0.55$.] Note that
our use of superpixels in the variable lightcurve detection mechanism
implies that our detection is nominally more sensitive to relatively
faint but large amplitude variables than bright but small amplitude
variables. Hence, we argue the pseudo-magnitude is a reasonably good
approximation for the real magnitude at maximum for our sample. In
addition, the majority of the variables in our sample is believed to
be AGB/RGB variables such as Miras and semi-regular variables (see
below), and their amplitudes are known to be quite large (e.g., $\ga
2.5\ \mbox{mag}$ for Miras, Whitelock 1996). This also
supports the view that the pseudo-magnitude is in fact a reasonable
alternative for the magnitude of variable for most of our lightcurves.

\setcounter{figure}{14}
\begin{figure}
\vspace{0.49\hsize}\centering{\tt fig15.gif}\vspace{0.48\hsize}
\caption{The analogue of a colour-magnitude diagram for the $r$-band
selected variables. Green dots are for group 1, blue for group 2, cyan
for group 3, and red for group 4 with $P\le 0.5\ \mbox{yr}$, and
yellow for group 4 with $0.5\ \mbox{yr}<P\le 1\ \mbox{yr}$.}
\label{fig:cmd}
\end{figure}

Each superpixel contains many variables and so it might be thought
that the period and pseudo-magnitude derived from the lightcurves may
be misleading. We expect this not to be the case because it is very
rare for two comparably bright variables to contribute to the same
superpixel. This can be seen on computing the mean distance between
our detected variables, which is $\sim$50 pixels or $\sim$7
superpixels.

The analysis of the spatial distribution of variables is based on the
catalogue of objects selected by Lomb's periodogram as having a
significant coherent behaviour. Datapoints lying more than 4 standard
deviations from the arithmetic mean are discarded to minimise the
effects of outliers in the periodogram. Given any sampling, there is a
correspondence between the peaks of the power spectrum and the
significance level. Specifically, we insist that the strongest
periodic signal has a false alarm probability of less than $10^{-7}$.
This effectively removes all the spurious lightcurves -- as can be
seen in Figure~\ref{fig:fakes}, in which the variables that
pass this criterion are shown as crosses. Additionally, we require that
the period be less than 800 days, roughly corresponding to the
baseline of the experiment. Only lightcurves with $18<r(\Delta f)\le
23$ are included. The rationale for these limits is that $r(\Delta f)=
23$ roughly corresponds to our detection limit, while variations with
$r(\Delta f)<18$ are most likely foreground contaminants. This gives a
variable star catalogue with 38779 members, which still has a number
of duplicate entries. These are removed using the following algorithm.
First, all the pairs of variable objects within 6 pixels are located,
and their lightcurves correlated. If the correlation coefficient is
larger than 0.75, then the lightcurve with lower significance is
removed from the catalogue. This leaves 35414 variable objects. The
positions, periods and pseudo-magnitudes of these variables are
available as an accompanying electronic table.

Figure~\ref{fig:peramp} shows the locations of the variables in the
period-$r(\Delta f)$ plane. We can quite easily divide the objects in
this space into several distinct groups, loosely corresponding to some
familiar variable star classes. For comparison,
Figure~\ref{fig:varstar} shows the locations of some common classes of
variable stars in the similar period-$R(\Delta f)$ plane, drawn from
the General Catalogue of Variable Stars\footnote{\tt\scriptsize
http://www.sai.msu.su/groups/cluster/gcvs/gcvs}. Here, $R(\Delta f)$
is the $R$ band flux difference between the maximum and the minimum
converted into magnitudes by the similar manner. Here, the variable
stars primarily reside in the Large Magellanic Cloud, and so there is
a vertical offset corresponding to the differences in distance modulus
and extinction. Clearly picked out in both Figures~\ref{fig:peramp}
and \ref{fig:varstar} are sequences corresponding to the
period-luminosity relations of population I and II Cepheids. Both
semi-regular and Mira variables are clustered together at longer
periods than the Cepheids.

\begin{figure}
\includegraphics[width=\hsize]{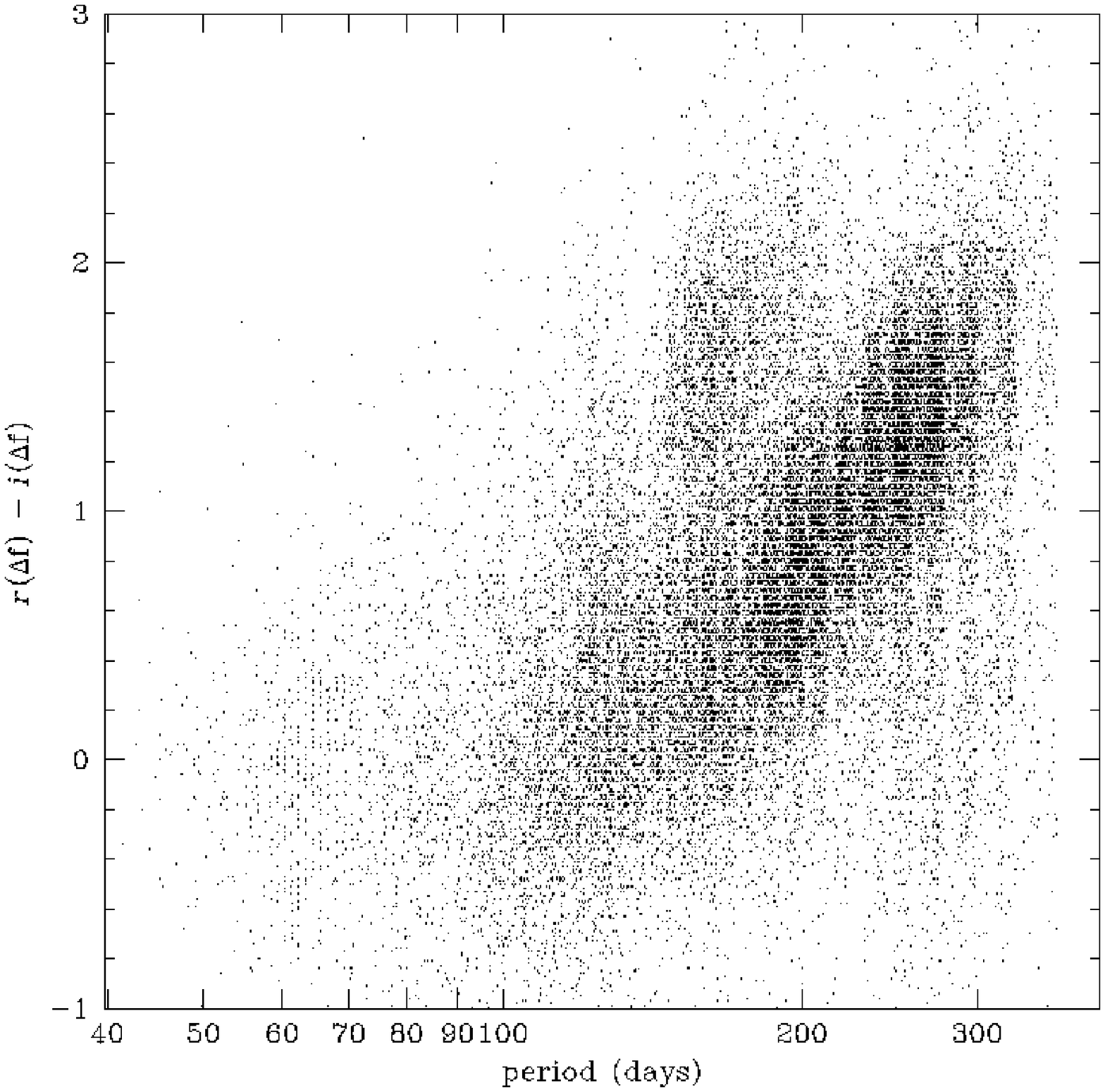}
\caption{Period versus colour $r(\Delta f)-i(\Delta f)$ for variables
in groups 3 and 4 with $P\le 1\ \mbox{yr}$.}
\label{fig:pcd}
\end{figure}

This motivates us to group the variables in Figure~\ref{fig:peramp}
into the following four divisions:
\begin{enumerate}
\renewcommand{\theenumi}{(\arabic{enumi})}
\item $r(\Delta f)+3\log_{10}P\le 25$ and $\log_{10}P\le 2$,
\item $r(\Delta f)+3\log_{10}P>25$ and 
      $r(\Delta f)+6\log_{10}P\le 32.2$ and $\log_{10}P\le 2$,
\item $r(\Delta f)+6\log_{10}P> 32.2$ and $\log_{10}P\le 2$,
\item $2<\log_{10}P\le 2.9$.
\end{enumerate}
The period is measured in days. Comparison between
Figures~\ref{fig:peramp} and \ref{fig:varstar} suggests the following
rough identifications. The variables in group 1 are mainly population
I Cepheids pulsating in either the fundamental or the first overtone
modes. In fact, the tracks on which the variables in group 1 reside
are suggestive of the classical Cepheid period-luminosity
relationships. This group also includes a distribution of variables
with periods shorter than 3 days. These are either eclipsing binaries
or population II Cepheids pulsating in the first overtone mode,
sometimes called BL Her or CWB stars (Feast 1996). The variables in
group 2 also cluster on a period-luminosity relationship and are
mainly population II Cepheids in the fundamental mode (W Vir or CWA
stars). Group 3 is composed mainly of semi-regular and Mira variables
with rather short periods or incomplete sampling, while those in group
4 are a mixture of semi-regular and long period Mira variables.
Typical lightcurves of objects in all four groups are given in
Figures~\ref{fig:lcurves1}-\ref{fig:lcurves4}.

In addition to the 35414 variables in the catalogue with reasonably
well-determined pseudo-magnitude and period, there are further
variables for which the reported period is longer than 800 days. We
classify these as group 5. The determined period and pseudo-magnitude
are not reliable since a whole cycle is not covered by the
experiment. A number of these are associated with CCD defects and can
be rejected with a suitable mask. This leaves 2243 variables with
ill-determined or incomplete periods after removing duplicate entries.
Some examples of lightcurves in this group can be found in
Figure~\ref{fig:lcurves5} while the data are available as an
accompanying electronic table. Variables in this group are not
included in subsequent calculations of statistical quantities.

\vfill
\subsection{Colours}

\begin{figure}
\includegraphics[width=\hsize]{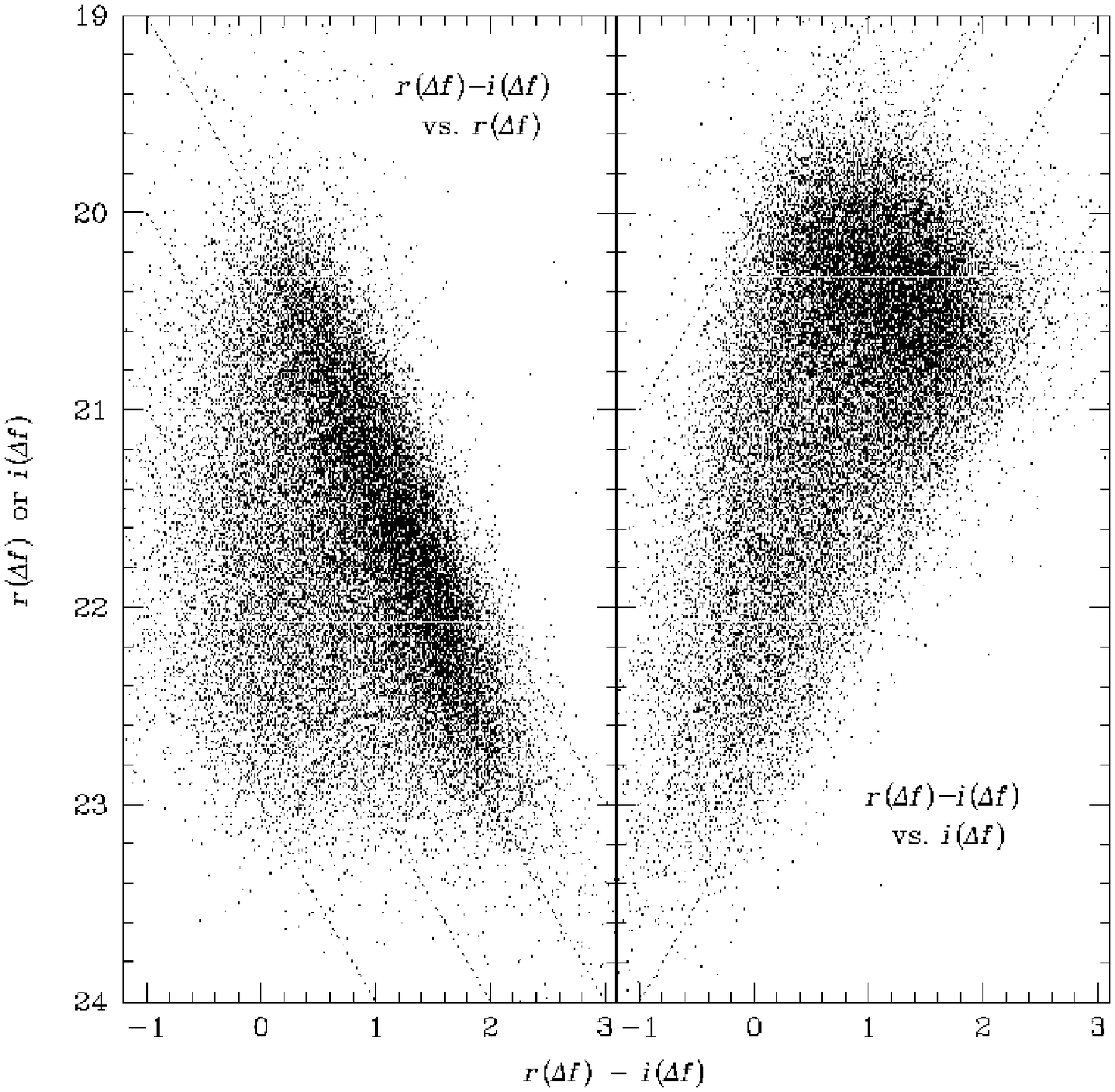}
\caption{Analogues of the colour-magnitude diagram for groups 3 and 4
with $P\le 1\ \mbox{yr}$.}
\label{fig:cmdlpv}
\end{figure}

For $r$-band selected variables, we examine the $g$ and $i$ band
lightcurves and, if they contain noticeable variations, we derive
pseudo-magnitudes $g(\Delta f)$ and $i (\Delta f)$. These can be used
to build an analogue of the colour-magnitude diagram, as shown in
Figure~\ref{fig:cmd}. Note that the $i$ band data are largely
restricted to the last two seasons, so variables with period longer
than 1 yr are not plotted.

\begin{figure*}
\includegraphics[width=15cm]{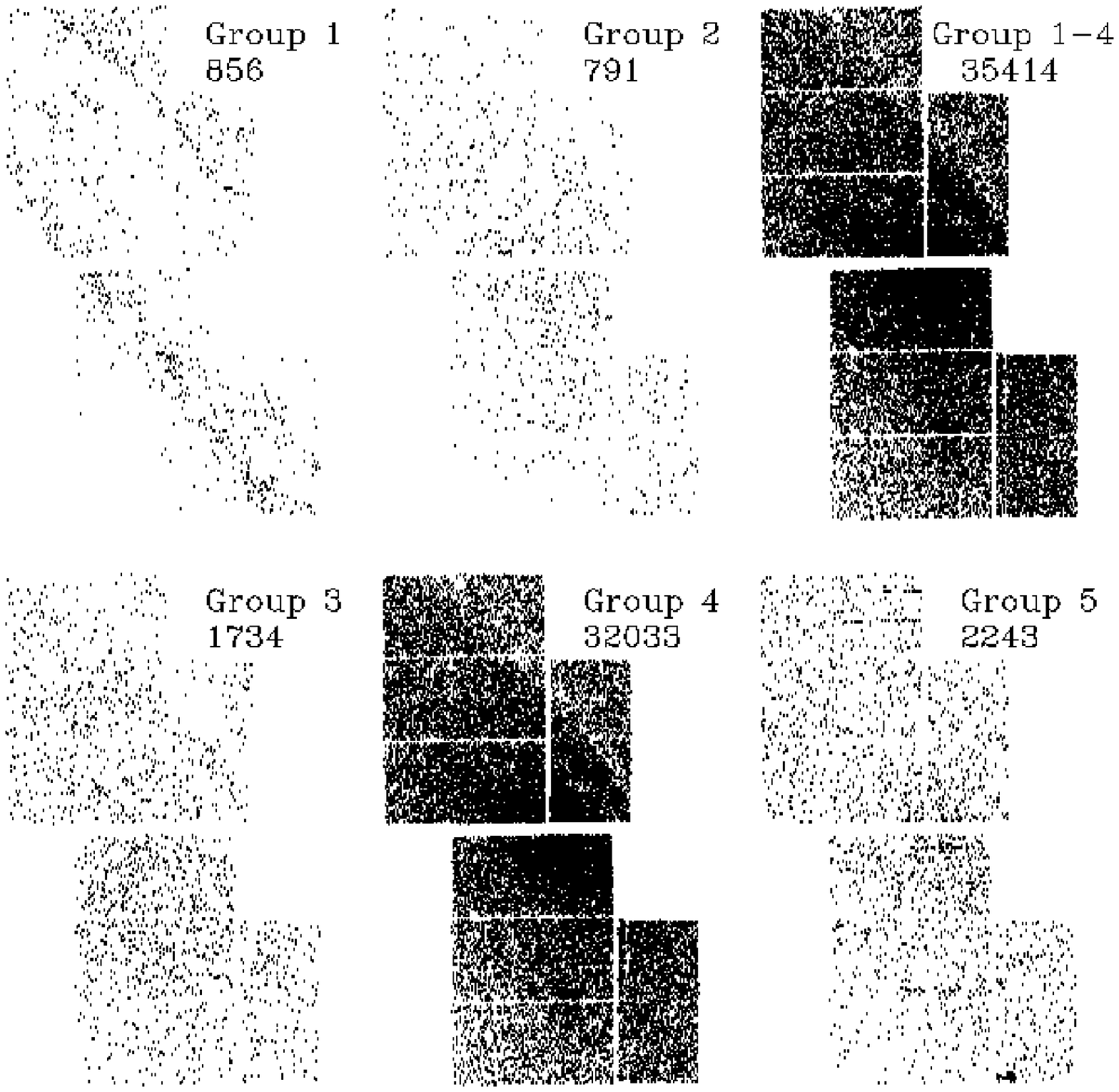}
\caption{The spatial positions of the variable stars, split into the
groups suggested by Fig.~\ref{fig:peramp}. The numbers of objects in
each group are shown in the top-right hand corner of each sub-panel.}
\label{fig:dist1}
\end{figure*}

The first thing to notice is that different groups of variable stars
are clearly separated. The variables in groups 1 and 2 are bluer than
those in groups 3 and 4. This is expected because Cepheids are F or G
(super)giants, whilst Miras are M (super)giants. As the variables in
group 1 are primarily population I Cepheids, their positions mark the
location of the instability strip. The positions of groups 3 and 4
indicate the location of the RGB and AGB of varying ages and
metallicity in this analogue of a colour-magnitude diagram. They are
segregated according to their periods, with the longer period
variables being brighter in $i$ and so redder in colours involving
$i$. However, this is not the case for the $g(\Delta f)-r(\Delta f)$
(not shown), as, along the sequence of M stars, the $r-i$ becomes
redder, whilst the $g-r$ stays the same (e.g., Margon et al.\ 2002;
Szkody et al.\ 2003). Figure~\ref{fig:pcd} shows the period plotted
against $r(\Delta f)-i(\Delta f)$ for groups 3 and 4 with $P\le 1\
\mbox{yr}$. This shows a moderately strong correlation, which is
analogous to the similar period-($H_p-K$) colour correlation for Mira
and semiregular variables (Whitelock, Marang, \& Feast 2000). Here,
$H_p$ is the Hipparcos broad-band magnitude. This
relation is also presumably related to the well-known period-$K$ band
luminosity relation for Mira variables (e.g., Feast et al.\ 1989; Noda
et al.\ 2002; Glass \& Evans 2003; Kiss \& Bedding 2003), as the
redder evolved AGB stars are typically brighter in near infrared
bands. The secondary clump centred on $\log_{10}P\approx 2.2$ and
$r(\Delta f)-i(\Delta f)\approx 1.8$ may be an artefact caused by
half-year sampling gaps combined with incompleteness of our period
determination, as it is displaced from the main correlation by almost
$\log_{\rm 10}2$. However, Whitelock et al.\ (2000) found a similar
``secondary clump'' in their AGB period-($H_p-K$) colour relation and
suggested that Miras with $P<225\ \mbox{days}$ are divided into two
groups depending on their $H_p-K$ colour or spectral types.
As the transformations
from our observables to these colours are unknown, the answer to
whether the secondary clump in Figure~\ref{fig:pcd} is real and indeed
an analogue to that found by Whitelock et al.\ (2000) is inconclusive
at the moment. Nevertheless, we note that $\sim\log_{\rm 10}2$
separation may also be explained physically by invoking pulsations on
the overtone mode.

Figure~\ref{fig:cmdlpv} shows analogues of the same $r(\Delta f)-
i(\Delta f)$ colour versus $i(\Delta f)$ magnitude diagram for only
groups 3 and 4 with $P\le 1\ \mbox{yr}$, along with the $r(\Delta f)-
i(\Delta f)$ colour versus $r(\Delta f)$ magnitude diagram. We note
that there is a strong cut-off in the bright end, namely $i(\Delta f)
\approx 20$ and this presumably corresponds to the tips of the AGB and
RGB with different ages and metallicities. The locuses are most likely
to be caused by the differences in metallicity and ages of the AGB and
RGB populations, with the redder side younger and more metal-rich. We
note that, along the locus, as the stars get redder, they become
fainter in $r$, but maintain more or less the same $i$
(psuedo-)magnitude.

\vfill
\subsection{Spatial Distribution}

\begin{figure*}
\includegraphics[width=15cm]{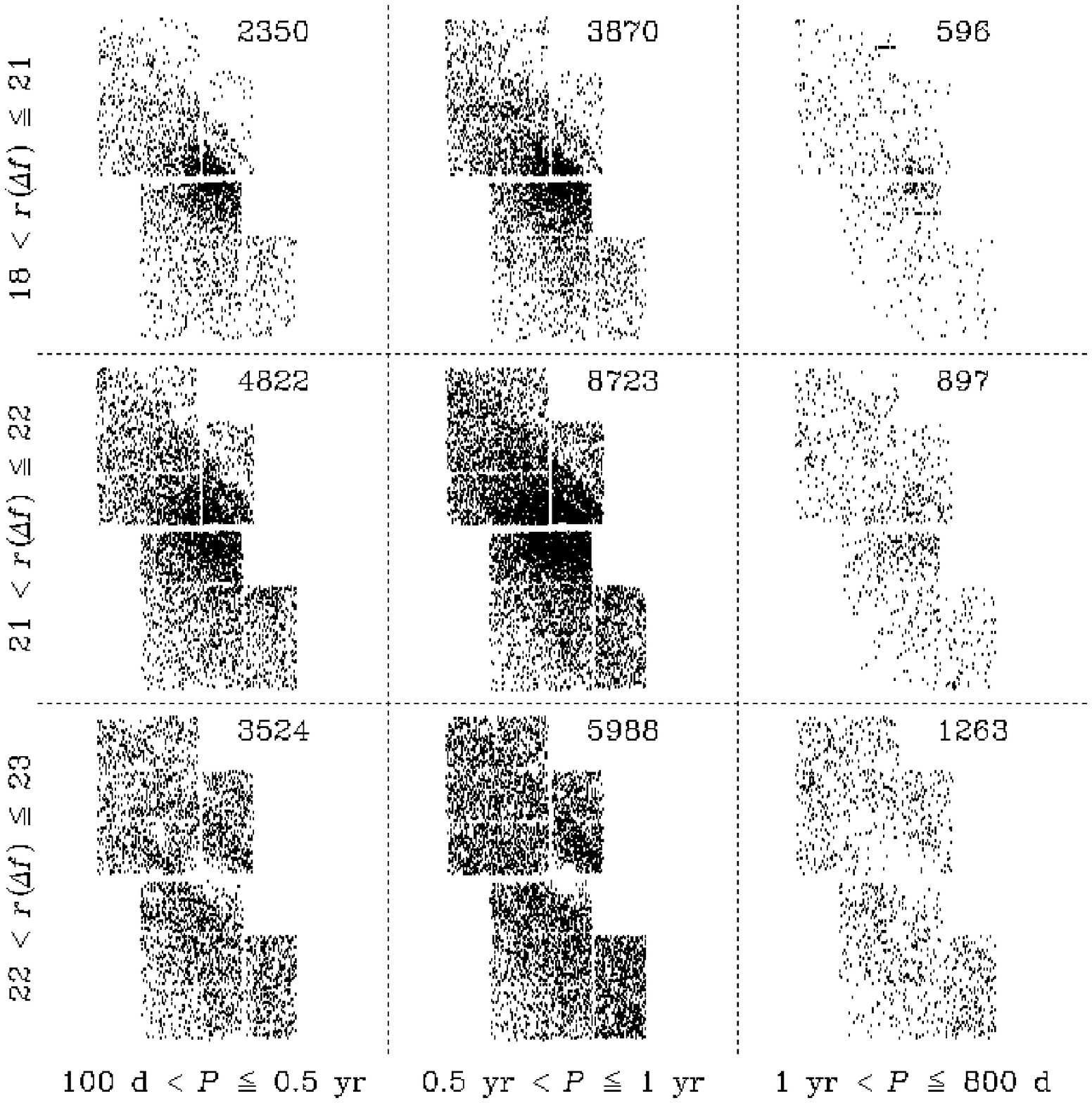}
\caption{The spatial positions of all the variables in group 4
(primarily Miras and long-period variables), subdivided according to
period and pseudo-magnitude. Note that there are obvious trends in
that the brighter variables or ones with shorter periods are more
centrally concentrated. The numbers of objects in each class are shown
in the top-right hand corner of each sub-panel.}
\label{fig:dist2}
\end{figure*}

Next, we investigate the spatial distribution of the groups of
variable stars on the WFC fields. Figure~\ref{fig:dist1} shows the
distributions of groups 1 to 5 of variable stars, together with the
set of all the 35414 variable stars with well-determined periods
(groups 1-4 combined). Figure~\ref{fig:dist2} shows the distributions
of variables in group 4, further sub-divided by their period and
pseudo-magnitude.

Most obviously, variables in group 1 follow the same pattern as the
bright supergiant population already seen in the top rightmost panel
of Figure~\ref{fig:distr}. Both populations seem to be associated with
the spiral or ring-like structure. This reinforces our identification
of group 1 as population I Cepheids, because the latter are known to
occur preferentially in spiral arms (Magnier et al.\ 1997; Mel'Nik,
Dambis \& Rastorguev 1999). Any spirality is much less evident in
groups 2, 3 and 4, as would be expected for population II Cepheids,
semi-regular variables and Miras, which are found in both the halo and
the old disc populations of M31 (Hodge 1992).

\begin{figure}
\includegraphics[width=\hsize]{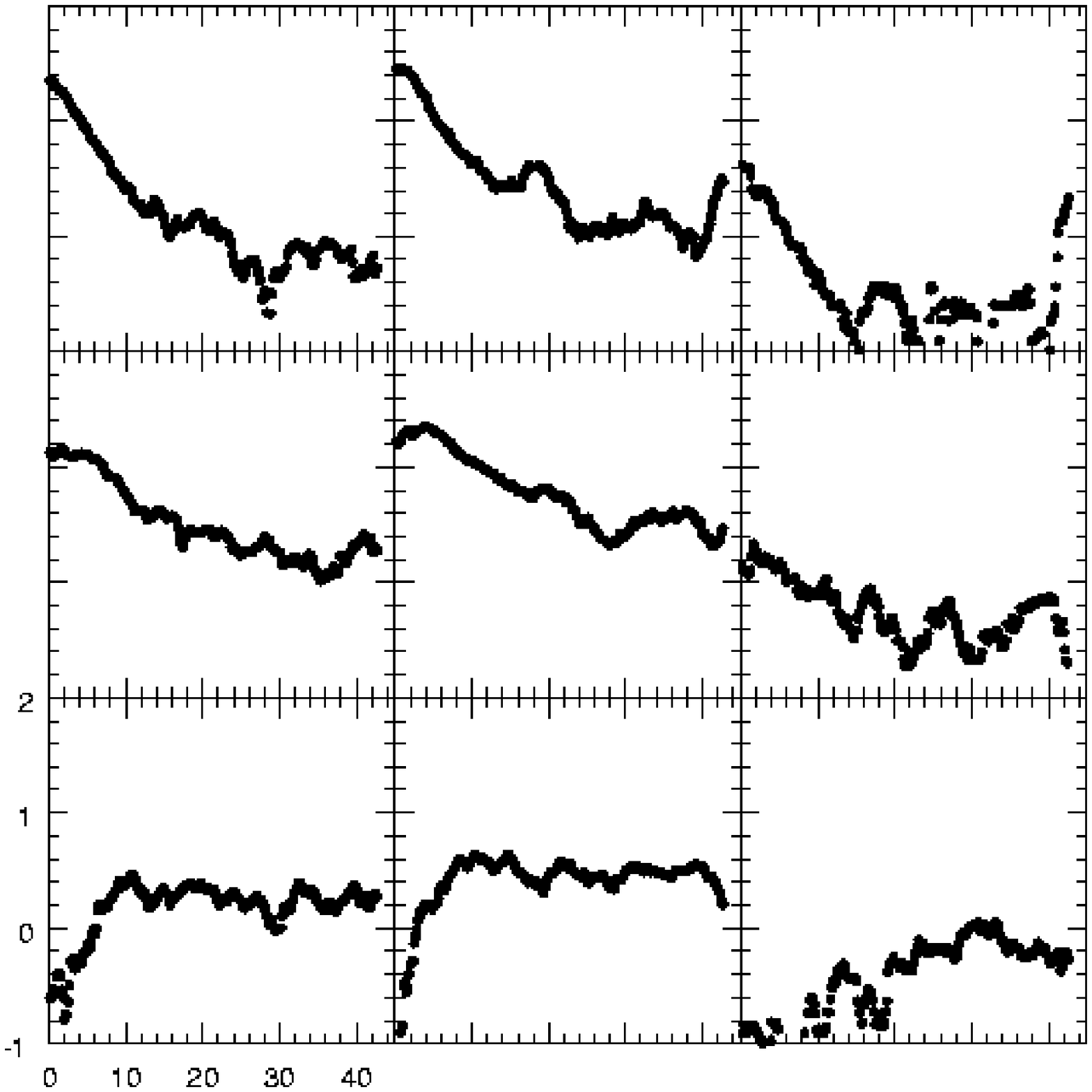}
\caption{The logarithm of the surface density plotted against distance
along the major axis towards the north-east for the classes of
variable object in Figure~\ref{fig:dist2}.}
\label{fig:surfprofiles}
\end{figure}

The variables in group 4 follow the same pattern as the AGB stars
already seen in the lower three panels of Figure~\ref{fig:distr}.
This reinforces the identification of group 4 as predominantly
semi-regular or Mira variables, as these correspond to pulsating AGB
and RGB stars. One can discern a number of trends in
Figure~\ref{fig:dist2}. First, for all but the faintest and longest
period variables, there is an overall depression of objects in the
north-west, which coincides with the prominent dust lane visible in
the colour map of Figure~\ref{fig:image}. Second, the central
concentration of objects varies noticeably, with the brighter (and
probably shorter period) variables being more centrally concentrated.
While the depression of detected variables with $r(\Delta f)>22$ near
the bulge is probably caused by the decline of detection efficiency
due to high surface brightness, one can still observe the clear trend
of a flattening in the number density gradient as the variation
becomes fainter (and longer). This is also illustrated in
Figure~\ref{fig:surfprofiles}, which shows the logarithm of the
surface density plotted against distance along the major axis for the
subdivisions of group 4 variables. The period of a Mira is believed to
be an indicator of the population to which it belongs (Whitelock
1996). Miras with periods less than 200 days are thought to be
primarily denizens of the halo, whilst those of longer period are more
massive and/or more metal-rich. The longer period Miras -- being more
massive and/or more metal-rich -- are therefore associated with
younger populations. In other words, the spatial distribution of the
younger Miras is more extended than that of the older ones. The same
conclusion can be drawn from the variation of central concentration
according to the brightness. The AGB is redder and therefore optically
fainter for a metal-rich population. This is consistent with the idea
that the visually brighter Miras represent older populations. All this is
consistent with the fact that the disk scalelength of M31 is larger in
the blue than the red (e.g., Walterbos \& Kennicutt 1988). It is also
consistent with a division of M31 stars into two populations, younger
and bluer ones in the disk (Population I), older and redder ones in
the bulge and spheroid (Population II), as discovered by Baade (1944).

\begin{figure}
\includegraphics[width=\hsize]{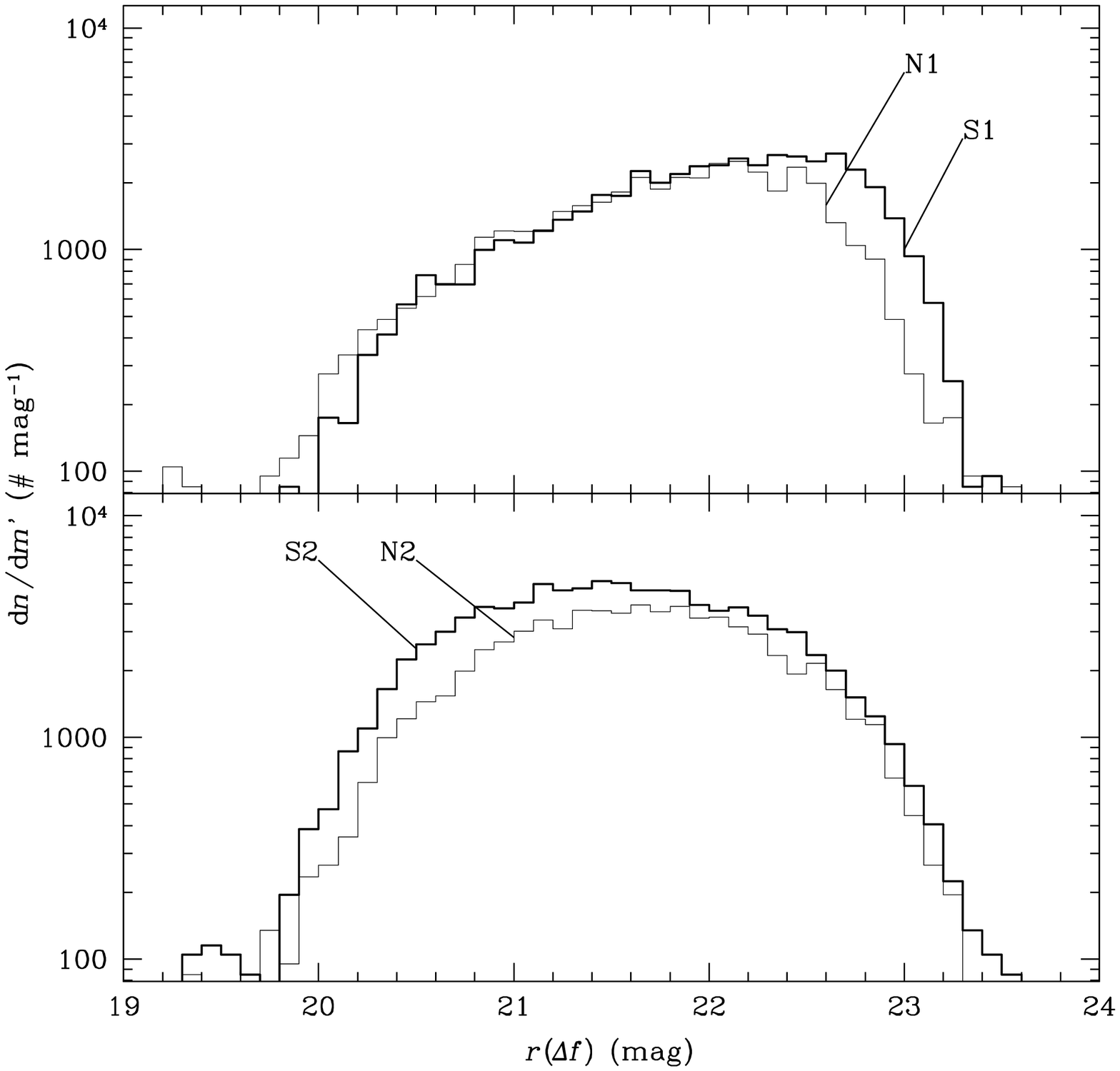}
\caption{The observed uncorrected luminosity function for the
variables stars in the same symmetrically positioned fields as
Fig.~\ref{fig:reshist}. The bin size is 0.1 magnitudes. Only the
variables with $P\le 800\ \mbox{days}$ are included.}
\label{fig:varhist1}
\end{figure}

\begin{figure}
\includegraphics[width=\hsize]{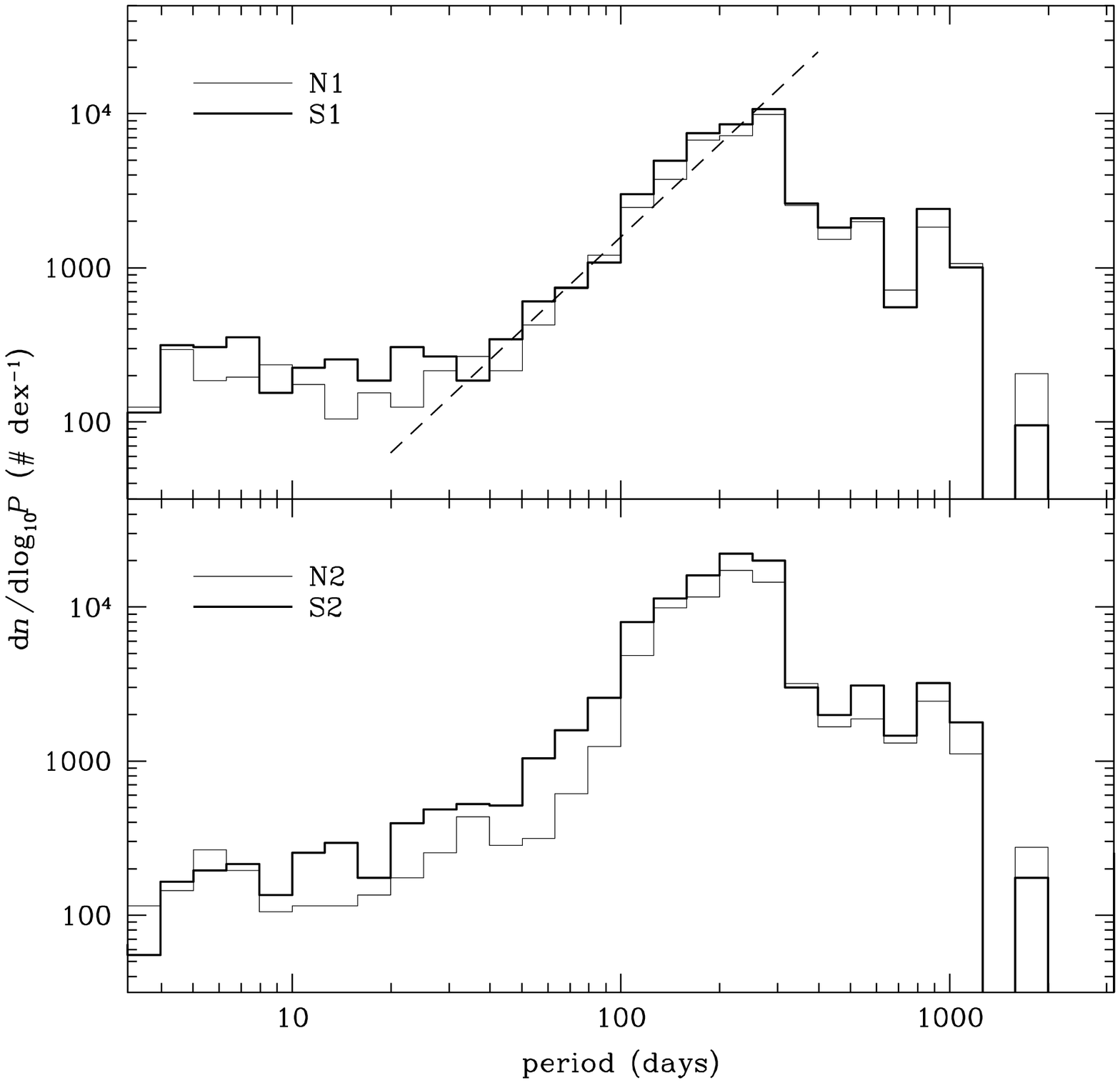}
\caption{The period distribution of the variable stars in the same
symmetrically positioned fields as Fig.~\ref{fig:reshist}. Only
variables with $18<r(\Delta f)\le 23$ are included. The dashed lines
has a slope of 2 and coincides with the rising part of the function in
all four fields.}
\label{fig:varhist2}
\end{figure}

We examine the distributions of variable stars broken down into
different pseudo-magnitude and period bins. For comparison purposes,
this is done for the same symmetrically positioned fields as used for
the resolved stars (N1, S1 and N2, S2 illustrated in the side-panel of
Fig.~\ref{fig:reshist}). Figure~\ref{fig:varhist1} shows the numbers
of variables with periods less than 800 days in each magnitude bin.
By a slight abuse of terminology, we refer to these distributions as
``luminosity functions''. For the exterior fields (N1 and S1), the
luminosity functions are in good agreement for $r(\Delta f)<22$. Note
that, even up to $r(\Delta f)=22$, the luminosity functions are still
rising. This suggests that incompleteness in our variable star
detection is not a substantial problem at least up to $r(\Delta f)=22$
for these exterior fields. For the interior fields (N2 and S2), there
is a difference in the overall normalisation causing the luminosity
functions to be offset. This is most likely due to the dust lane in
the N2 field. The luminosity functions for the interior fields rise
more steeply than for the exterior fields. This is probably caused by
underlying differences in the stellar populations, as the bulge
contribution dominates in the interior fields. The luminosity
functions for the interior fields also start to decline at $r(\Delta
f)=21.5$. This can be understood by the higher surface brightness in
the bulge, which in turn means that the detection efficiency is lower.
Figure~\ref{fig:varhist2} shows the numbers of variables with $18<
r(\Delta f)\le 23$ as a function of the logarithm of the period. The
bin size is 0.1 dex. For the exterior fields, there is a good
agreement between the two distributions. The distribution is flat
($dn/dP\propto P^{-1}$) between periods of 1 and 40 days, reflecting
predominantly the properties of the Cepheid population. Between 40
days and 0.5 yr, the population is made up mainly of semi-regular
variables and Miras and the distribution is well-approximated by a
power-law ($dn/dP\propto P$). The fall-off for longer periods may be
caused by the decline in detection efficiency because of incomplete
sampling. For the interior fields, there is a depression in the
distribution derived for the N2 field as compared to the S2 around
periods of 100 days. This in turn suggests that it is fewer objects
with $P\sim 100\ \mbox{days}$ that are responsible for the smaller
number of detected objects in the N2 field.

As for the resolved stars discussed in \S~\ref{sec:image}, we
construct smooth surface number density distributions of the variable
stars. Here, we use a window function that is a square box of size
3\arcmin. The result is compared with the surface brightness map and
the resolved star ($R\le 21$) density map in Figure~\ref{fig:images}.
{\it None} of the three images show any sign of CCD to CCD or field to
field overall scale variation. This is very reassuring as it suggests
that our variable star detection is not systematically biased. Second,
the resolved star distribution shows a much more prominent ring-like
structure (possibly just the spirality) than the variable star
distribution. Third, the projected shape of the triaxial bulge is much
less evident in the variable star map than in the other two maps.

\section{The Asymmetries of the Variable Stars}
\label{sec:asymmetry}

\setcounter{figure}{24}
\begin{figure}
\includegraphics[width=\hsize]{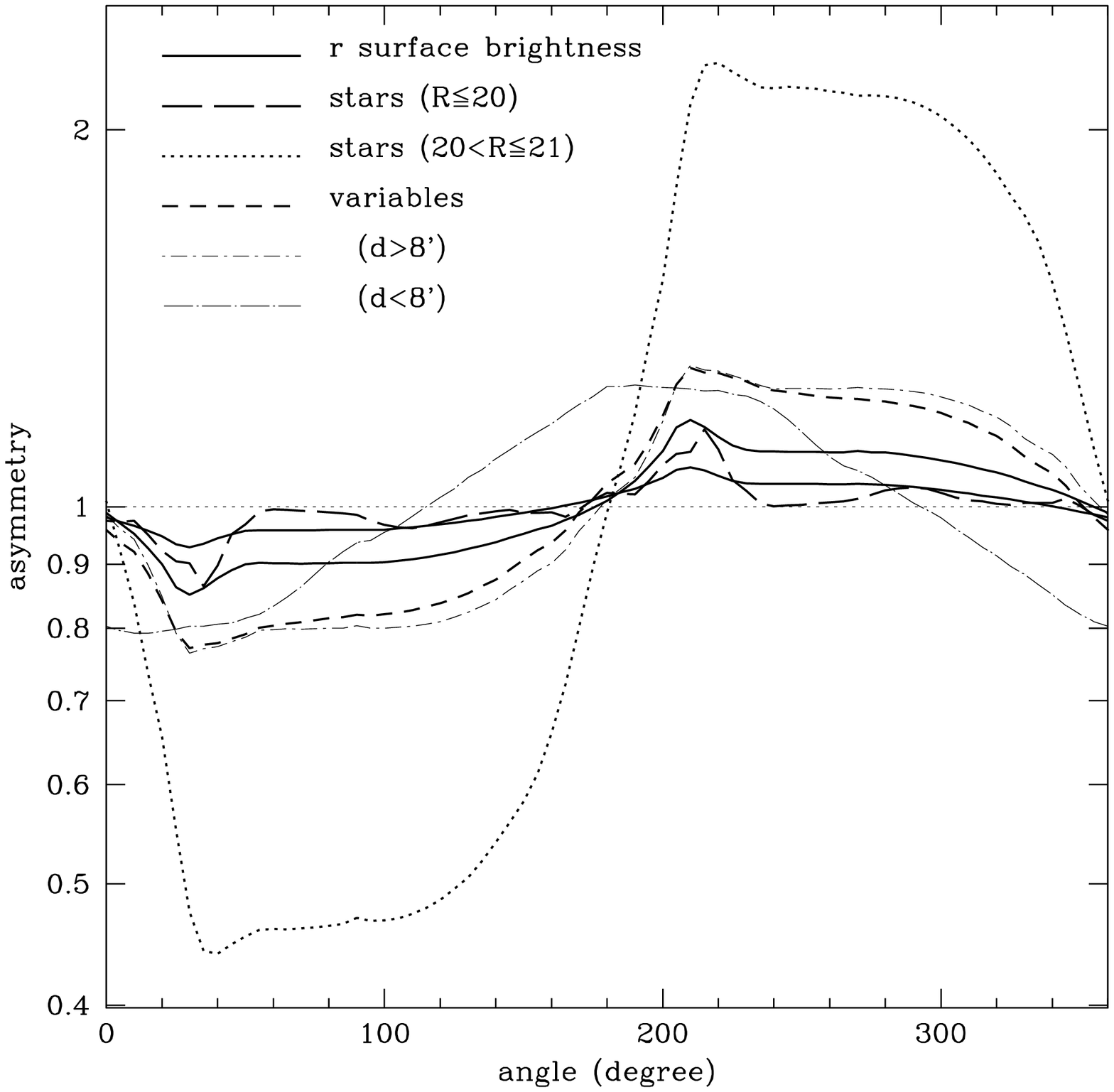}
\caption{The ratio of the mean density on either side of an axis
plotted against position angle of the axis for the brightest resolved
stars (long dashed line), intermediate brightness resolved stars
(dotted line), variable stars (short dashed line) and the integrated
$r$ band light (two solid lines -- see text for difference). Also
shown (the two dot-dashed lines) are the variables divided according
to projected distance to the centre of M31.}
\label{fig:ratios}
\end{figure}

To examine the asymmetry of these distributions, we construct division
images -- that is, each image is divided by a copy of the image which
is rotated by 180\degr\ with respect to the M31 centre. The resulting
division maps are displayed in Figure~\ref{fig:diffimages}. In all
three cases, it is clear that the far side (south-east) is brighter or
has more detected objects than the near side (north-west). More
importantly, the asymmetry pattern of all three images quite closely
follow each other, with the dust lane in the north-west clearly
visible in all cases. It is therefore clear that a crucial assumption
underpinning the microlensing experiments towards M31 -- namely, that
the variable star population is unlikely to show an asymmetry between
the near and far side (Crotts 1992) -- is unfortunately not correct.
The intrinsic distributions are probably well-mixed, but the effects
of differential extinction and the prominent dust lanes associated
with the spiral structure in M31 cause all the observed distributions,
whether of resolved stars or of variable stars, to be markedly
asymmetric.

\begin{figure}
\centering
\includegraphics[width=\hsize]{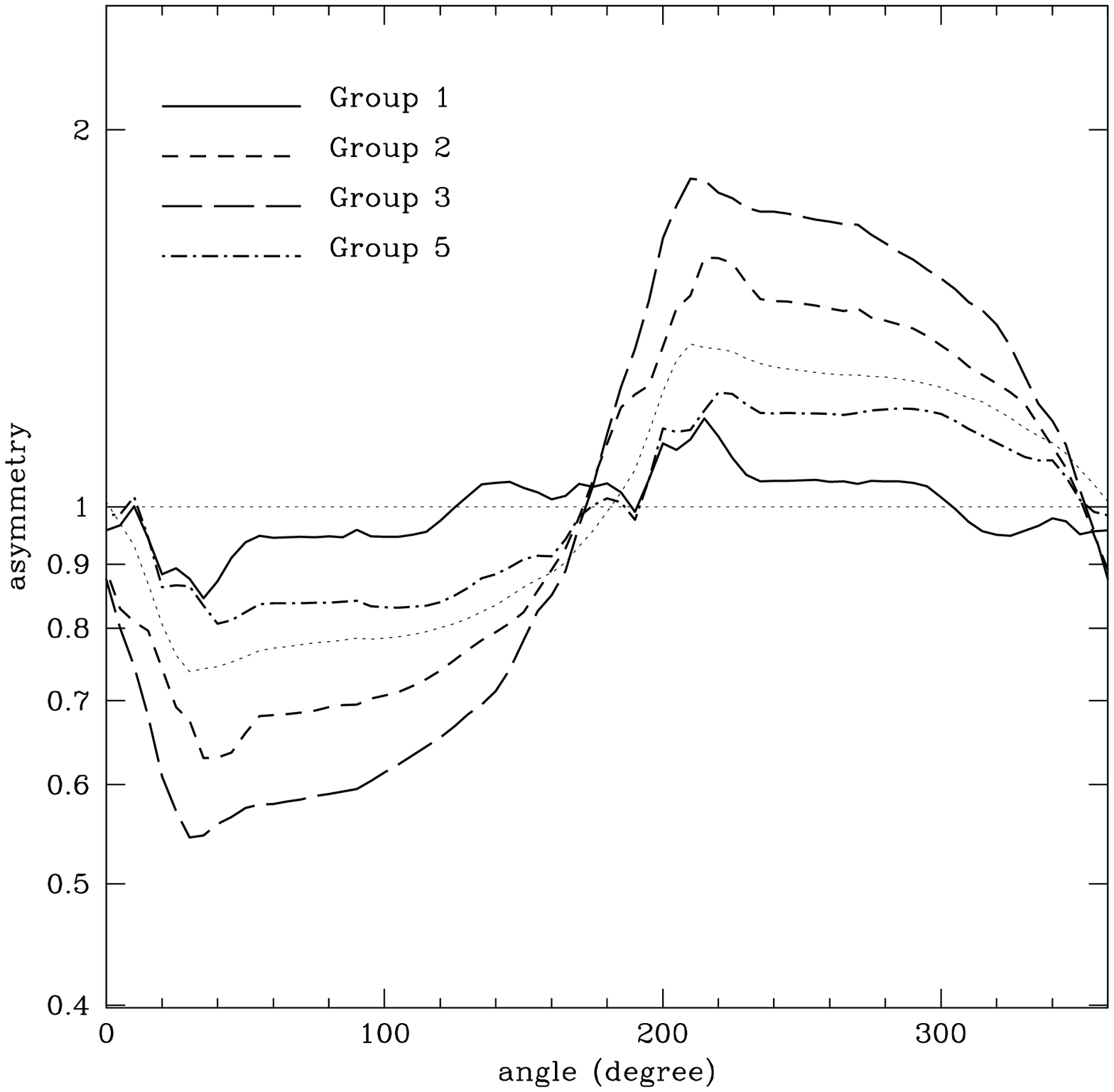}
\caption{The asymmetry signal as a function of position angle for the
variable stars in groups 1, 2, 3 and 5. The dotted curve shows the
asymmetry signal of the 35414 variable stars in the catalogue. Both
Fig.~\ref{fig:asone} and Fig.~\ref{fig:astwo} are not corrected for
the effects of M32 or HD 3914, but we have shown in the text that the
correction is small.}
\label{fig:asone}
\end{figure}

Any axis through the centre divides the galaxy into two parts. The
ratio of the mean density on each side changes as the axis changes. We
call the direction in which this ratio is maximised the asymmetry
axis. To find the asymmetry axis, we construct the logarithm map of
the division image and compute the ratio of the mean density of each
side as a function of the position angle of the axis, measured
anti-clockwise from north through east. The results are shown in
Figure~\ref{fig:ratios}. The long-dashed line is for the brightest
resolved star ($R\le 20$) density, while the dotted line is for the
intermediate brightness resolved star ($20<R\le 21$) density. The two
solid lines are derived from the $r$ band image. The curve showing the
larger asymmetry amplitude assumes the minimum contribution from the
sky, while the smaller amplitude curve assumes the maximum
contribution, and thus the reality should be somewhere in between. The
maximum contribution is inferred from the reduced images without
sky-subtraction, while the minimum contribution is derived from the
south-east corner of field 2, CCD 1, which is assumed to contain no
contribution from M31. Finally, the short-dashed line shows the
asymmetry of the variables. This can be divided into the contributions
from the bulge and disc respectively, by crudely dividing into two
samples according to the projected distance $d$. The asymmetry of the
whole variable sample follows that of the disc variables alone. This
is because the asymmetry signal is weighted according to projected
surface area, and so the effect of the bulge is minimal. Note that the
contribution from M32 and a $\sim$7th magnitude star (HD 3914) lying
just at the southern edge of the southern field CCD 2 is corrected for
by excising a comparable area (7\arcmin$\times$7\arcmin\ mask) in both
fields. In any case, their effects are minimal, again because they
occupy very little of the total area.

The asymmetry of the brightest resolved stars -- most likely young,
massive supergiants -- is virtually negligible. The asymmetry of the
resolved stars with $20<R\le 21$ is quite noticeably enhanced compared
to the total surface brightness. Some of this may be caused by the
varying detection efficiency of the CCDs which is related to the
choice of different reference images, as discussed in
\S~\ref{sec:image}. However, the fact that the pattern of the curve
follows that for the surface brightness suggests that the effect is
real, even if the magnitude is exaggerated. Such artifical enhancement
is not expected for the variable stars, as the choice of reference
image plays a less crucial role in the variable star detection
algorithm. The resolved stars are searched for on different reference
images, which can have different photometric conditions, whereas the
variables are searched for using the entire lightcurves and so the
photometric conditions average out. In addition, both the magnitude of
the asymmetry and the shape of the curve emphasise the fact that the
variable star asymmetry is clearly associated with the underlying
surface brightness. The maximum asymmetry occurs around $\sim$40\degr,
while the symmetric division is around $\sim$180\degr, which is more
or less an east/west division. For comparison, the position angle of
the major axis of the outer disk is $\sim$38\degr\ (de Vaucouleurs
1958), while the position angle of the bar is between 45\degr\ and
50\degr\ (Hodge \& Kennicutt 1982). The fact that the maximum
asymmetry occurs when the division lies almost parallel with the major
axis provides further evidence that the asymmetry is mainly caused by
the dust lane, as this too lies almost parallel with the major axis
(see Fig.~\ref{fig:image}).

\begin{figure}
\centering
\includegraphics[width=\hsize]{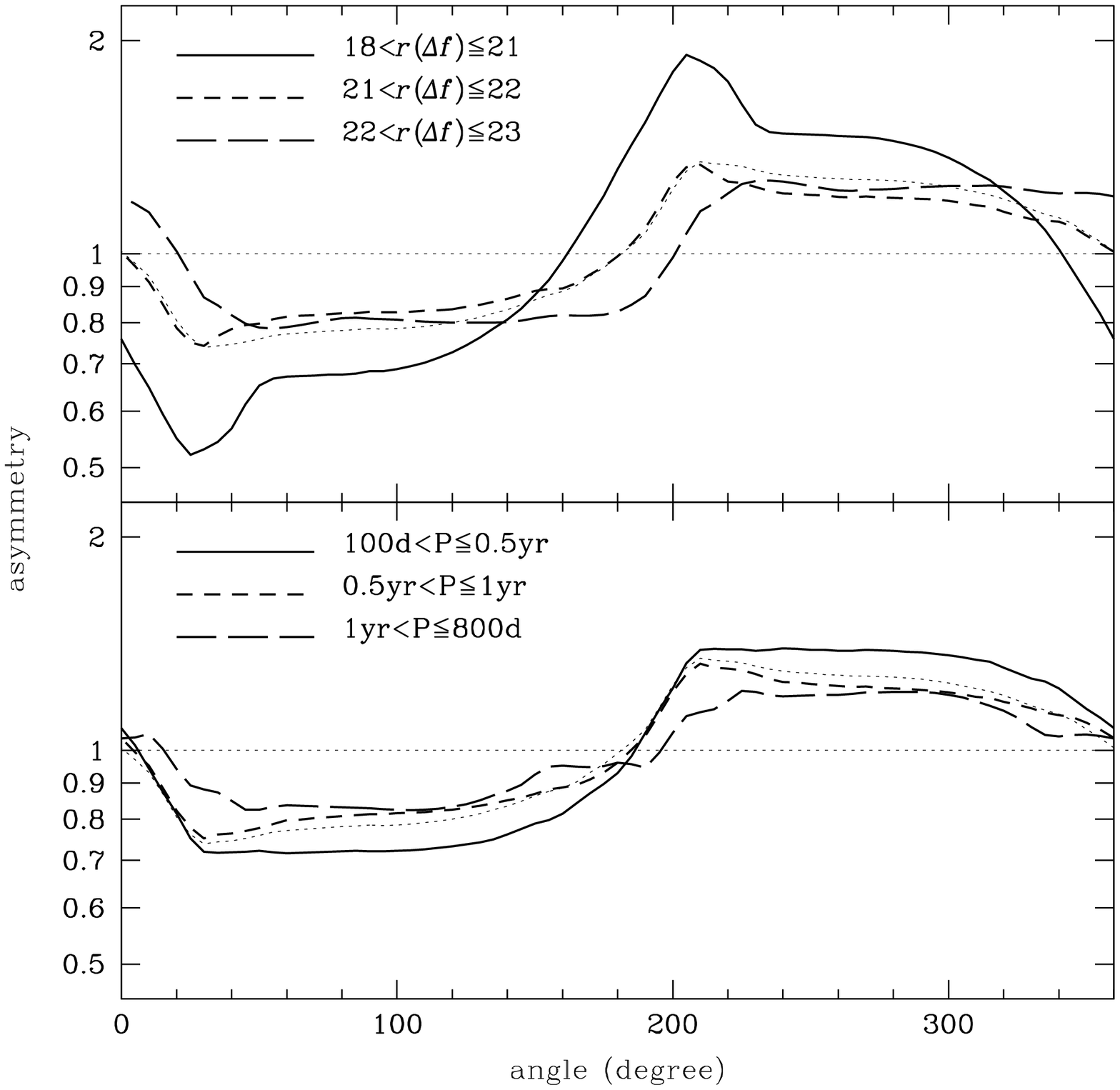}
\caption{The asymmetry signal as a function of position angle for the
variable stars in class 4, broken up according to pseudo-magnitude
(top) and period (bottom). The criteria for the divisions of
pseudo-magnitude and period are the same as recorded in
Fig.~\ref{fig:dist2}. Also plotted in the dotted curve is the
asymmetry signal of the 35414 variable stars in the catalogue.}
\label{fig:astwo}
\end{figure}

The asymmetry of the variable stars according to their group, period
and pseudo-magnitude can be studied by exactly the same method. The
results are shown in Figures~\ref{fig:asone} and \ref{fig:astwo}.
Since the effects of M32 and HD 3914 are not corrected for, the
asymmetry is a bit larger than for the entire set of 35414 variables.
We notice that the asymmetry signal of variables in group 1
(predominantly Population I Cepheids) is similar to that of the
supergiant stars in Figure~\ref{fig:ratios}. Also, we deduce that the
variable star asymmetry follows the asymmetry of the long period
variables with $100\ \mbox{days}<P\le 1\ \mbox{yr}$ and with $21<
r(\Delta f)\le 22$. These are about $\sim$40\% of the variable star
catalogue. There is a mild suggestion from the Figures that the
asymmetry is larger for brighter variables, which is consistent with
the enhanced asymmetry of the resolved star density between $20<R\le
21$. Probably both of these populations correspond to the brightest
end of the AGB.

\vfill
\section{Correlations with Other Catalogues}

\begin{figure}
\includegraphics[width=\hsize]{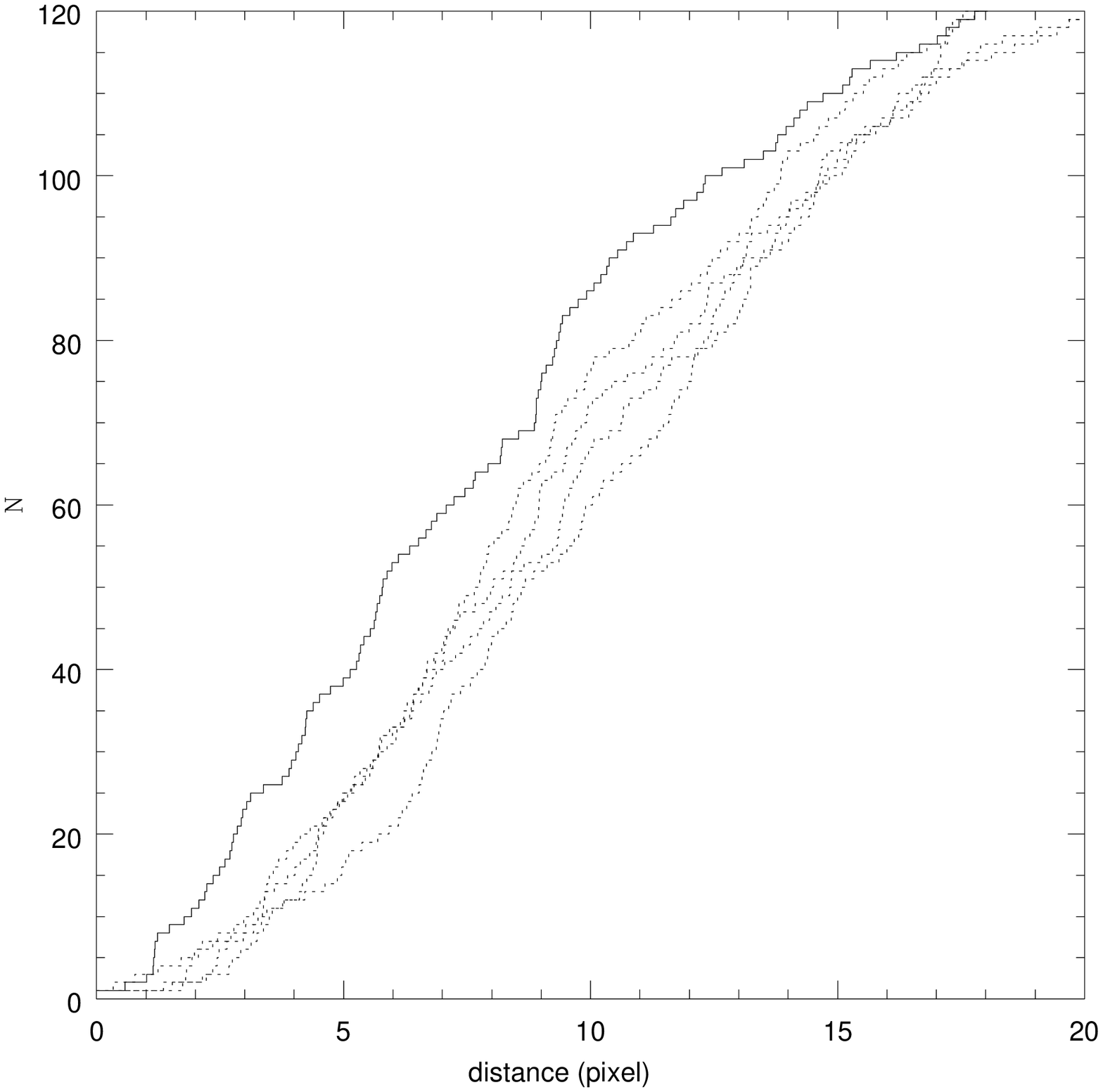}
\caption{The full line shows the cumulative number of pairs matched
between the optical and X-ray variable source catalogues as a function
of separation. The dotted lines are constructed by first translating
the optical catalogue by prescribed amounts in different directions
and then correlating. So, this gives an idea of the number of
accidental alignments expected.}
\label{fig:alignments}
\end{figure}

It is useful to compare our catalogue of variable stars with other
surveys and archives of M31. Here, we correlate the catalogue with the
discrete X-ray sources and the known novae.

\begin{table*}
\begin{minipage}{150mm}

\caption{Known novae detected in our catalogue. The number listed as
the PA lightcurve refers to the number for the corresponding
POINT-AGAPE superpixel lightcurve, which is also given in the
subpanels in Fig.~\ref{fig:novalcurves}. For some cases, more than one
nova-like superpixel lightcurve around a single nova position are
detected as separate variable stars in our catalogue. However, all
three cases of double detections show additionally variability in the
baseline of the second lightcurve given in parentheses. The positions
are directly measured from the resolved images of the novae near the
maximum in the WFC $r$ frames.}
\label{table:novae}

\begin{tabular}{lrccccl}
\hline Nova $^\rmn a$ &Alert&\multicolumn{2}{c}{Position (J2000.0)}
&PA lightcurve& Identifier$^\rmn b$&Reference\\ &&RA&Dec&&\\
\hline
EQ~J004249+411506 $^\rmn {c, d}$
&Jul 1999&00\h42\m49\fs7&+41\degr15\arcmin08\arcsec&78668&\ldots&1\\
EQ~J004222+411323 $^\rmn e$
&Aug 1999&00\h42\m22\fs3&+41\degr13\arcmin26\arcsec&83479&\ldots&2\\
EQ~J004241+411911
&Aug 1999&00\h42\m41\fs1&+41\degr19\arcmin11\arcsec&25851&PACN-99-05&3\\
EQ~J004244+411757 $^\rmn f$
&Jul 2000&00\h42\m44\fs0&+41\degr17\arcmin57\arcsec&26946&PACN-00-01&4\\
EQ~J004237.6+411737 &Aug 2000
&00\h42\m37\fs7&+41\degr17\arcmin39\arcsec&26021 (25568)&PACN-00-04&4\\
EQ~J004257+410715
&Oct 2000&00\h42\m57\fs1&+41\degr07\arcmin17\arcsec&77716&PACN-00-06&5\\
EQ~J0042.5+4114 $^\rmn d$
&Jul 2001&00\h42\m30\fs8&+41\degr14\arcmin36\arcsec&81539&PACN-01-01&6\\
EQ~J0042+4112
&Aug 2001&00\h42\m18\fs5&+41\degr12\arcmin40\arcsec&83835&PACN-01-02&7,8\\
{[OBT2001]~3} $^\rmn g$ &Sep 2001
&00\h42\m34\fs6&+41\degr18\arcmin13\arcsec&26277 (25695)&\ldots&9\\
EQ~J004303.3+411211.3
&Oct 2001&00\h43\m03\fs3&+41\degr12\arcmin12\arcsec&77324&PACN-01-06&10\\
EQ~J004233+411823 $^\rmn h$ &Jan 2002
&00\h42\m33\fs9&+41\degr18\arcmin24\arcsec&26285 (26121)&\ldots&11\\
EQ~J004252+411510 $^\rmn h$ &Jan
2002&00\h42\m52\fs9&+41\degr15\arcmin11\arcsec&79136&\ldots&11\smallskip\\
($\stackrel{?}{=}$ CXOM31~J004318.5+410950)$^\rmn i$ 
&(Jul/Aug 2001)&00\h43\m18\fs6&+41\degr09\arcmin50\arcsec&74935&\ldots&\\
\hline
\end{tabular}

\medskip
$^\rmn a$ The name of the nova follows SIMBAD.\\
$^\rmn b$ The identifier shows the label given by Darnley et al.\
(2004) to the nova.\\
$^\rmn c$ The alert occurs before our first observation. However, the
lightcurve shows the behaviour consistent with a fading nova. This is
also studied in \citet{Ri01}.\\
$^\rmn d$ The lightcurves are quite noisy.\\
$^\rmn e$ $\stackrel{?}{=}$ 
C84~25 / CXOM31~J004222.3+411333 / CXOM31~J004222.4+411334
\citep{Ka02,Ko02}.\\
$^\rmn f$ $\stackrel{?}{=}$ CXOM31~J004243.9+411755 \citep{Ka02}.\\
$^\rmn g$ $\stackrel{?}{=}$ XMMU~J004234.1+411808 / CXOM31~J004234.4+411809
\citep{Os01,TBP01,Ko02}.\\
$^\rmn h$ The alert occurs near the very end of our observation. Only
sudden brightening at the end is observed in our lightcurves.\\
$^\rmn i$ We find the variable with a nova-like lightcurve within
0\farcs8 of CXOM31~J004318.5+410950 \citep{Ka02}. No alert for this
nova has been reported during its outburst.\smallskip\\
References. --
(1) \citealt{ML99}; (2) \citealt{JML99}; (3) \citealt{Li99};
(4) \citealt{LPA00}; (5) \citealt{Do00}; (6) \citealt{Li01a};
(7) \citealt{ML01}; (8) \citealt{As02}; (9) \citealt{FMC01};
(10) \citealt{Li01b}; (11) \citealt{Fi02}.

\end{minipage}
\end{table*}

\subsection{X-ray Catalogues}

M31 has recently been surveyed by both the XMM-Newton and Chandra
satellites (Osborne et al.\ 2001; Kong et al.\ 2002). In particular,
Chandra surveyed the inner 17\arcmin$\times$17\arcmin\ and found 204
discrete X-ray sources. Our WFC fields are separated by a gap of
$\sim$2\arcmin, and so the very centre of M31 is not covered. This gap
is rich in discrete X-ray sources, and so only $\sim$140 of the
objects listed in Kong et al.\ (2002) lie within our WFC fields.

Figure~\ref{fig:alignments} correlates the variable sources in our
catalogue with the discrete X-ray sources. To allow for possible
identification with poorly sampled or sporadic optical sources, the
primitive version of the catalogue with 97280 lightcurves is used.
The figure shows the cumulative distribution of the separation
distances. Of course, some accidental alignments are expected. To
estimate how serious this problem is, we also show curves of the
cumulative distributions with the optical positions shifted by 42
pixels north-east, north-west, south-east and south-west
(Hornschemeier et al.\ 2001). This corresponds to a shift of 30 pixels
in each compass direction. The claimed accuracy of the astrometry in
both our and Kong et al.'s (2002) catalogue is 1\arcsec. This suggests
that objects can be identified if their positions coincide within an
error circle of 1\farcs4 or $\sim$5 pixels. Judging from
Figure~\ref{fig:alignments}, the number of accidental alignments is
$\sim$25, whereas the actual number of correlations is $\sim$40, which
suggests that $15\pm 8$ identifications are physically meaningful.

Some of the identifications are due to globular clusters. Whilst some
of the globular clusters are indeed optically varying, the core of an
unresolved globular cluster may also be misreported as a resolved star
and this may cause spurious lightcurves in the catalogue. A handful of
the identifications may be due to cataclysmic variables and soft X-ray
transients. These are binary systems with a compact degenerate object
and a low-mass star. Typically, they are undetectable in quiescence
($V=28-30$) but brighten dramatically in outbursts ($V=20-22$) and so
may then be present in the survey.

Fourteen of the 40 correlations coincide with the positions of
semi-regular variables, although for two of these instances, our
sampling does not allow us to rule out the possibility of a bursting
rather than pulsating star. Semi-regular variables are not expected to
emit X-rays, unless they are the binary companion to a compact
object. Hence, some of these alignments may well correspond to
cataclysmic variables. Lightcurves for 6 of these objects are shown in
Figure~\ref{fig:xraystars}. We note that the lightcurves of r3-43 and
r3-46 (in the notation of Kong et al.\ 2001) have an X-ray outburst
$\sim 50$ days before the optical maximum. A further two objects,
r3-87 and r3-48, coincide with optical variables showing a linear fall
in the first two seasons. One of these is the lightcurve labelled
19755 shown in the top-left panel of Figure~\ref{fig:lcurves5}.

\vfill
\subsection{Nova Alerts}

Nova alerts are published regularly in {\it IAU Circulars}. There are
14 nova alerts in M31 during the period that spans our WFC
observations. We look for detections in our primitive variable star
catalogue within a 3\arcsec\ error circle of the reported nova
position. This yields 12 matches to nova-like lightcurves. Both
missing cases, EQ~J004249+411632 (Filippenko et al.\ 1999) and
EQ~J004241+411624 (Fiaschi et al.\ 2002), lie very close to the centre
of M31 and fall in the gap between the two fields.
Table~\ref{table:novae} gives the position, time of burst and
reference of the 12 candidates, whilst the lightcurves are shown in
Figure~\ref{fig:novalcurves}. All 12 detections are located within a
few arcminutes of the centre of M31 and are either of fast or
moderately fast speed class, presumably reflecting the experimental
setup of the observers who issued the {\it IAU Circulars}. In
addition, we find one further nova (CXOM31~J004318.4+410950) by
correlating with the X-ray catalogue of Kaaret (2002). We note that
another nova (EQ~J004244+411757) also lies within 1\farcs3 of the
position of an X-ray point source in the same catalogue. The X-ray
counterparts of both novae have been claimed as super-soft X-ray
sources by Di Stefano et al.\ (2003). A more detailed and homogeneous
selection of novae using POINT-AGAPE data is performed in Darnley et
al.\ (2004), who find 7 of the novae in Table~\ref{table:novae}.
Darnley et al.\ (2004) also identify 13 other novae present in M31
further from the bulge and with longer durations and slower speed
classes.

\setcounter{figure}{30}
\begin{figure}
\vspace{0.49\hsize}\centering{\tt 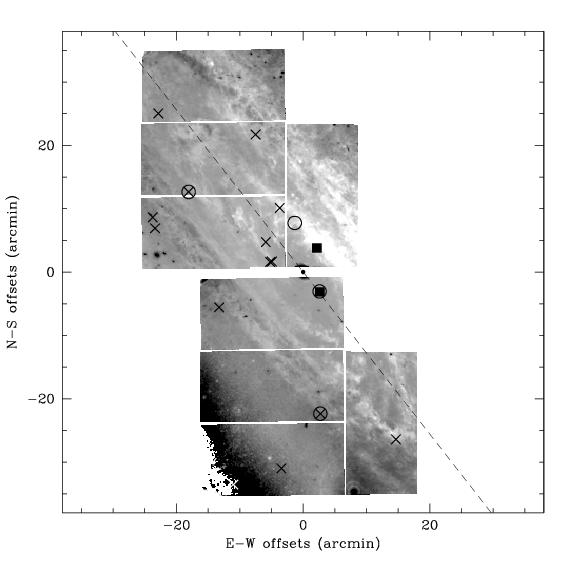}\vspace{0.48\hsize}
\caption{Locations of claimed microlensing events in M31, overplotted
on the $g-r$ colour map. Circles show POINT-AGAPE candidates, crosses
MEGA candidates and squares WeCAPP candidates. Also shown by a dot is
the centre of M31, while the dashed line lies at position angle
38\degr\ (measured North through East) and marks the major axis of the
outer disk.}
\label{fig:candidates}
\end{figure}

\vfill
\section{Discussion}

Figure~\ref{fig:candidates} shows the locations of 17 reported
microlensing candidates (taken from Paulin-Henriksson et al.\ 2003,
Riffeser et al.\ 2003 and de Jong et al.\ 2004) overplotted on the
$g-r$ colour map. There is a clear asymmetry, as de Jong et al.\ point
out. However, the origin of the asymmetry is difficult to say, as
there are no events where the galaxy is dusty and a preponderance of
events in the bright spiral arms. The early papers on microlensing
towards M31 made the assumption that variable star populations would
show no signs of asymmetry between fields north and south of the major
axis. Microlensing, on the other hand, will show a substantial
asymmetry if the lensing population resides in a spheroidal halo.
Early on, it therefore seemed that an asymmetry in the locations of a
sample of microlensing events would be enough to identify the lensing
population conclusively with the dark halo of M31. The work in this
paper has demonstrated that this is not the case. All the variable
star distributions are asymmetric in the sense that the far side (or
south-east) is brighter or has more detected objects than the near
side (or north-west). This is caused primarily by the effects of
variable extinction and the prominent dust lanes associated with the
spiral structure. Therefore, the near-far asymmetry signal of
microlensing is unfortunately mimicked in the variable star
populations, in the underlying surface brightness and in the resolved
star catalogue. This makes an accurate measurement of the asymmetry
signal caused by microlensing more difficult. This paper has measured
the size and direction of the asymmetry signal as a function of the
period and pseudo-magnitude of the variable stars. These data form the
zero-point of the asymmetry scale. Microlensing by a dark halo is a
possible explanation if the sample of microlensing candidates of given
amplitude and period is more asymmetric than the corresponding
variable stars.

Theoretical calculations of the expected magnitude of the near-far
asymmetry signal from microlensing (e.g., Kerins et al.\ 2001, 2003)
assume that the source population shows no preference between the near
and far side. However, as the resolved star distribution is also
asymmetric, it is clear that this assumption is questionable as well.
The source population for the identifiable microlensing events is
typically bright (though unresolved) stars, and so it too is likely to
be asymmetric. Hence, the magnitude of the near-far asymmetry signal
as calculated theoretically needs to be amplified by an additional
factor. The values of the asymmetry reported in Kerins et al.\ (2003)
must now be taken as lower limits.

For a spherically symmetric halo, Kerins et al.\ (2003) found that the
underlying pixel-lensing rate in the far-disk of M31 is already more
than 5 times the rate of near-disk events. Clearly, a still more
substantial asymmetry is expected when the effects reported in this
paper are included. However, successful measurement of the component
of any asymmetry signal caused by microlensing is now much more
difficult. In principle, the additional effects can be modelled and
corrected for, but in practice this will make the results less robust.
One way to overcome the difficulties is to limit the study to
microlensing events of sources only brighter than some fixed magnitude
at baseline. Analysis of the microlensing asymmetry as a function of
the source flux will however require a large sample of events with
high signal-to-noise ratio.

Another way of overcoming the difficulties is to look for an east-west
asymmetry rather than a near-far asymmetry. Both the resolved stars
and the various groups of variables are more-or-less symmetric about
an axis running north-south, as seen in Figures~\ref{fig:ratios} and
\ref{fig:asone}. So, this suggests that the underlying populations of
variables are indeed well-mixed and that their asymmetry has a common
origin in the dust distribution and possibly some intrinsic disk
surface density asymmetry. If so, we can correct for it by choosing
the north-south axis as our reference axis. Even though the east-west
microlensing asymmetry is less than the near-far asymmetry in the
absence of dust problems, it is still asymmetric, as the eastern side
contains much of the far disk. This effect should be detectable in
cases when the halos contains a high fraction of massive compact
objects.

\section{Conclusions}

The POINT-AGAPE collaboration has constructed a catalogue of the
locations, periods and brigtness of 35414 variable stars in M31. This
is a by-product of our microlensing search. The variable stars have
been classed into four groups -- roughly corresponding to population I
and II Cepheids, semi-regular and Mira variables of short period and
long period Miras. The spatial distribution of the variable stars
shows a number of interesting trends. In particular, the brighter and
shorter period Miras are more centrally concentrated than their
fainter and longer period counterparts. This suggests that
the younger populations of AGB stars are more extended than the older
one. The variable star catalogue has been correlated with the X-ray
point source catalogue available from surveys with the Chandra
satellite. After taking into account accidental alignments, we find
that there are $\sim$15 physically meaningful identifications. This
finding will lead to additional complications in the measurement of
any asymmetry due to microlensing. In particular, it may be
advantageous to look for an east-west asymmetry rather than a near-far
asymmetry.

\vfill
\section*{Acknowledgements}
The data were taken through the Isaac Newton Group's Wide Field Camera
Survey Programme. This research has made used of the SIMBAD database
operated at Centre de Donn\'{e}es astronomiques de Strasbourg,
Strasbourg, France. We thank Albert~K.~H.~Kong for providing us with
the unpublished lightcurves of selected X-ray point sources in M31 and
Vasily Belokurov for guidance on variable star catalogues. Work by JA
and YT has been supported through a grant from the Leverhume Trust
Foundation. NWE thanks the Royal Society (UK) and the Particle Physics
and Astronomy Research Council (UK) for support. SCN was supported by
the Swiss National Science Foundation and by the Tomalla Foundation.
AG was supported by grant AST~02-01266 from the National Science
Foundation (US).

%\vfill

\vspace{\vsize}

\setcounter{figure}{9}
\begin{figure}
\caption{({\tt fig10.gif})
Lightcurves of variable stars in group 1, folded by their
periods. For visual aid, two cycles are plotted. The superpixel flux
in ADU s$^{-1}$ is plotted against time in days. The number in the
right-hand corner is the identifier in our catalogue. Lightcurve 29439
is most likely an eclipsing binary, whilst the remaining lightcurves
are Population I Cepheids. Of these, 19897 and 20539 have the nearly
symmetric shape associated with the first overtone, whilst the
remainder are pulsating in the fundamental mode.}
\label{fig:lcurves1}
\end{figure}
\begin{figure}
\caption{({\tt fig11.gif})
Folded lightcurves of variable stars in group 2.
\hspace{\hsize}}\label{fig:lcurves2}
\end{figure}
\begin{figure}
\caption{({\tt fig12.gif})
Folded lightcurves of variable stars in group 3.
\hspace{\hsize}}\label{fig:lcurves3}
\end{figure}
\begin{figure}
\caption{({\tt fig13.gif})
Folded lightcurves of variable stars in group 4.
\hspace{\hsize}}\label{fig:lcurves4}
\end{figure}
\begin{figure}
\caption{({\tt fig14.gif})
Lightcurves of variable stars with ill-determined or
incomplete period (group 5).
\label{fig:lcurves5}}
\end{figure}

\setcounter{figure}{22}
\begin{figure}
\caption{({\tt 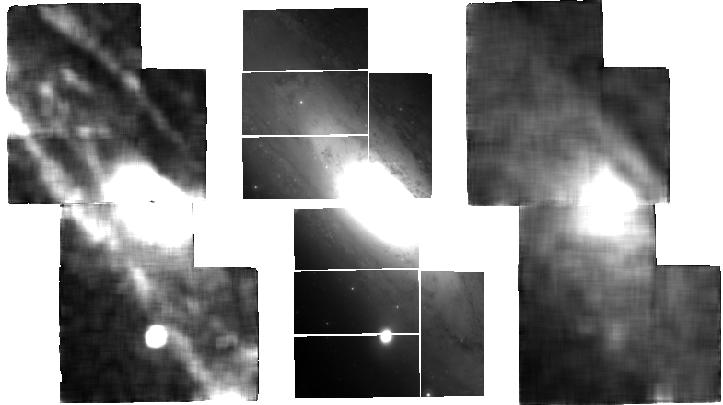})
Grey-scale surface density maps of the resolved stars with $R
\le 21$ (left) and variable stars (right), together with the surface
brightness in the $r$ band (centre). Notice the ring-like structure
visible in the resolved star map.}
\label{fig:images}
\end{figure}
\begin{figure}
\caption{({\tt 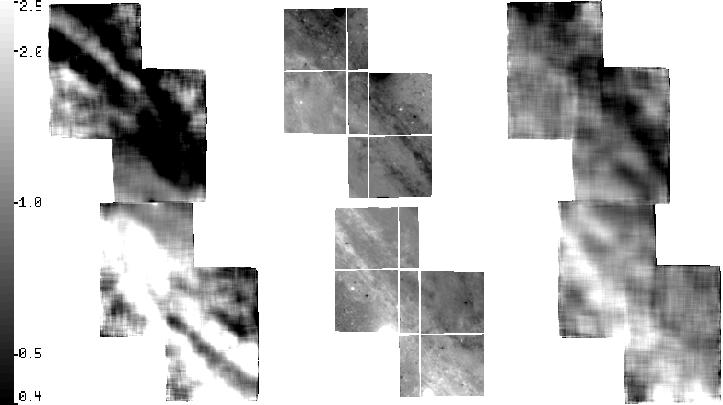})
Grey-scale division maps of the resolved stars (left) and
variable stars (right), together with the surface brightness (centre).
The surface density is divided by its 180\degr\ rotated image to form
the division map, so the map gives a direct measure of the asymmetry.}
\label{fig:diffimages}
\end{figure}

\setcounter{figure}{28}
\begin{figure}
\caption{({\tt fig29.gif})
Optical lightcurves of the possible matches with discrete
X-ray sources in Kong et al.\ (2001). The numbers in the top left of
each panel refers to the possible match and the separation.}
\label{fig:xraystars}
\end{figure}
\begin{figure}
\caption{({\tt fig30.gif})
POINT-AGAPE lightcurves of known novae. The flux in ADU
s$^{-1}$ is plotted against time in JD-2451000. The number in the
right-hand corner is the identifier in our catalogue, also listed in
Table~\ref{table:novae}.}
\label{fig:novalcurves}
\end{figure}

\end{document}